\documentclass[english,tightenlines,eqsecnum,amsmath,amssymb,aps,prd,superscriptaddress,nofootinbib,showpacs,floats]{article}
\usepackage{amsmath,amsfonts,amssymb,multicol}
\usepackage{graphicx,caption,subcaption}
\usepackage{cite}
\usepackage[usenames]{xcolor}
\definecolor{AHZ}{rgb}{0.0,1,0.0}
\usepackage[colorlinks,linkcolor=AHZ,citecolor=red,bookmarks]{hyperref}
\usepackage{rotating}
\usepackage[mathscr]{eucal}
\usepackage{graphicx}
\usepackage[T1]{fontenc}
\usepackage{babel}
\usepackage{authblk}
\usepackage{color}
\usepackage{amssymb}
\usepackage{fancyhdr}

\def\nn{\nonumber\\}
\newcommand{\f}[2]{\frac{#1}{#2}}
\def\be{\begin{equation}}
\def\ee{\end{equation}}
\def\bea{\begin{eqnarray}}
\def\eea{\end{eqnarray}}

\def\bwt{\begin{widetext}}
	\def\ewt{\end{widetext}}

\usepackage{amsmath}

\usepackage{amssymb}

\begin{document}
	
\title{Casimir Traversable Wormholes in Gauss-Bonnet Gravity}
\author{${^1}$Mohammad Reza Mehdizadeh\thanks{mehdizadeh.mr@uk.ac.ir}~}
\author{~${^2}$Amir Hadi Ziaie\thanks{ah.ziaie@riaam.ac.ir}}
\affil{{\rm ${^1}$Department~of~ Physics,~ Faculty~ of~Science,~Shahid~ Bahonar~ University of Kerman, P.~ O.~ Box~ 76175, Kerman, Iran}}
\affil{{\rm ${^2}$Research~Institute~for~Astronomy~and~Astrophysics~of~ Maragha~(RIAAM), University of Maragheh,  P.~O.~Box~55136-553,~Maragheh, Iran}}
\renewcommand\Authands{ and }
\maketitle
\begin{abstract}
In this work, we study Casimir wormhole solutions and investigate satisfaction of energy conditions in the framework of Einstein-Gauss-Bonnet (EGB) gravity. We firstly find traversable wormhole solutions supported by a general form for the Casimir energy density. Applying a specific redshift function, we choose suitable parameters that respect the energy conditions in the vicinity of the throat due to the presence of higher order curvature terms. Then, we check the energy conditions considering non-constant redshift function and for suitable choice of model parameters, we find that these conditions are satisfied at infinity as well as in the vicinity of the throat. The stability of the Casimir traversable wormhole solutions are investigated using the Tolman-Oppenheimer-Volkoff (TOV) equation. Finally, we study trajectory of null as well as timelike particles in the wormhole spacetime.
\end{abstract}
\maketitle
\section{Introduction}
As cosmic shortcuts connecting distant regions of spacetime, wormholes are known as theoretical solutions to the equations of General Relativity (GR) that allow for the possibility of long journeys across the Universe~\cite{mt}. First formulated in the early 20th century, GR describes how mass and energy shape the spacetime curvature, providing then the best known description for gravitational interactions. The history of wormhole solutions in GR dates back to 1916, when the Austrian physicist Ludwig Flamm looking over Schwarzschild's solution, noticed that another solution was also possible which described a hypothetical time-reversed black hole later known as a {\it white hole}~\cite{Flamm}. However, it was not until 1935 that Albert Einstein and Nathan Rosen formalized the concept of a bridge connecting two regions of spacetime, now known as the Einstein-Rosen bridge~\cite{ERosen1935}. These early solutions were later realized non-traversable and representing mathematical curiosities rather than physical realities. The modern era of wormhole physics began in the 1980s with the pioneering work of Morris, Thorne, and Yurtsever, who proposed the first traversable wormhole solutions in the context of GR~\cite{mt}. Their work demonstrated that wormhole configurations could, in principle, be stabilized and made traversable if exotic matter with negative energy density were used to keep the wormhole throat open and support it against gravitational collapse~\cite{mt,mt1,mviss}. This groundbreaking research laid the foundation for subsequent studies on the issues of traversability, stability, dynamics, and potential applications of wormholes and paved the way to investigate such problems in alternative theories of gravity.
\par
A critical aspect of wormhole physics {in GR} is the role of exotic matter in ensuring the traversability and stability of these spacetime structures. Exotic matter is characterized by violation of the null energy condition (NEC) that the existence of which is required to prevent the wormhole throat from collapsing under its own gravity~\cite{mviss}. The necessity of exotic matter poses significant challenges, as it is not observed in classical physics and is inconsistent with the energy conditions typically satisfied by ordinary matter distribution. This requirement has fueled research into potential sources of exotic matter, including phenomena from quantum field theory such as Casimir effect~\cite{caseffect,casieffect} and Hawking evaporation process~\cite{Ford1988}. While the negative energy sources provide an interesting and tempting test-bed to study traversable wormholes, the violation of energy conditions, at least in classical GR, is unacceptable from the general viewpoint of physicists. This {issue} has led researchers to explore alternative mechanisms, such as modified theories of gravity in which, the extra degrees of freedom or higher-order curvature correction terms ({not present in GR}) can effectively mimic the properties of exotic matter, possibly help avoiding or at least minimizing the need for exotic matter which is crucial for stability and traversability of a wormhole configuration. Several investigations have been carried out along this line of research among which we can quote, traversable wormholes in higher-dimensional theories~\cite{needexmodgr}, brane world scenarios~\cite{needexmodgr1}, Lovelock~\cite{needexmodgr2} and $f(R)$ gravity theories~\cite{needexmodgr3}, scalar-tensor theories~\cite{needexmodgr4s}, Einstein-Cartan theory~\cite{needexmodgr5}, $f(R,T)$~\cite{needexmodgr6} and teleparallel gravities~\cite{teleworm}, Rastall theory~\cite{needexmodgr7} and other modified gravity theories~\cite{needexmodgr8}.
\par
Modified theories of gravity have received a considerable amount of attention in recent decades as alternatives to GR, particularly in addressing cosmological as well as astrophysical issues such as dark energy, dark matter, and the inflationary phase of the early Universe~\cite{modgracosas}. Among these {theories}, EGB gravity has emerged as a particularly promising framework which extends GR by incorporating higher-order curvature terms that become significant in high-energy regimes, such as those near the Planck scale or in the early universe~\cite{EGBmod},\cite{EGBmod1}. EGB gravity is also of interest because it avoids some of the pathologies associated with other modified gravity theories, such as ghost instabilities, making this theory a robust candidate for exploring the phenomena that result from gravitational interactions~\cite{modgracosas},\cite{ghostegb}. The inclusion of EGB correction terms in the gravitational sector of the action leads to novel nontrivial implications in cosmological~\cite{EGBmod2} and astrophysical scenarios~\cite{EGBmod3}, see also~\cite{EGBmod1} and references therein. The study of wormhole structures in EGB gravity has garnered significant attention during the last years since, the presence of higher-order curvature terms can act as an effective source of exotic matter, potentially eliminating the need for usage of exotic matter to make the wormhole stable and traversable~\cite{needexmodgr4},\cite{msmf}. These new insights into the stability, traversability, geometry and potential observational signatures of wormholes {provide} a rich framework for further research in EGB gravity. 
\par  
As stated above, an important and open challenge in wormhole physics is to search for viable matter sources that can support the wormhole geometry. A potential candidate for such matter sources is the Casimir effect which generates negative energy densities between closely spaced conducting plates. The {study} of wormhole structures in the presence of Casimir effect was triggered by the works~\cite{mt1},\cite{mviss},\cite{trigcasworm},\cite{trigcasworm1},\cite{trigcasworm2}. {The first two works have discussed the use of the Casimir energy to support a wormhole geometry. In~\cite{trigcasworm}, the authors have explored Casimir effect for a quantized massive scalar field with non-conformal coupling in a wormhole spacetime. The author of~\cite{trigcasworm1} obtained the Casimir energy-momentum tensor induced in the vacuum state of the massless conformally-coupled scalar field by the geometry and topology of a long wormhole throat. He further examined whether this exotic matter is sufficient to stabilize a macroscopic wormhole configuration. However, the term \lq{}\lq{}{\it Casimir Wormhole}\rq{}\rq{} was coined for the first time by Remo Garattini who tried to extend the studies performed in~\cite{mt1,mviss} towards utilizing Casimir effect as a possible source for a traversable wormhole~\cite{trigcasworm2}. Very recently, the author of this paper has also extended the usage of Casimir energy source to the cases of rotating~\cite{casiesrot} and charged rotating~\cite{ccasiesrot} wormhole configurations.} After these pioneering studies, wormhole configurations in such exotic spacetimes {have} been widely surveyed in the literature among which one may highlight, Casimir wormholes in modified gravity theories~\cite{caswomodgr}, effects of generalized uncertainty principle (GUP)~\cite{caswomodgup} and a Yukawa term~\cite{casyuk} on Casimir wormholes, Casimir wormholes in Yang-Mills theory~\cite{Yangcasworm} and other works~\cite{caseworm11}. { Moreover, the study of GUP-corrected wormhole geometries in the presence of Casimir effect between two parallel conducting plates in 5D EGB gravity has been performed in~\cite{zubire1} and the combined effects of a point charge’s electromagnetic field with the negative Casimir energy density have been discussed in~\cite{garatin2023,mfzh12}. Of particular interest are n-dimensional Casimir wormhole geometries in EGB gravity under the influence of thermal effects~\cite{garatini2025,Muniz:2025teu}, where the authors found that thermal corrections modify the wormhole's geometry through increasing spatial curvature in the throat region and thus affecting its traversability. In all of the above GB wormhole studies, the redshift function is considered to be zero. Here, we relax this assumption and consider a specific radial-dependent choice for the redshift function, which	tends to zero at spatial infinity. Indeed, since the GB term has low effects on regions far from throat, one can expect that a non-constant redshift function may contribute to solutions satisfying the energy conditions. Motivated by the above investigations, our aim in the present work is to build and study wormhole solutions in EGB gravity with general Casimir energy for various physical structures subject to different boundary conditions recently introduced in~\cite{Ziameh1}. We therefore perform our investigation in the presence of $r$-dependent redshift function and examine the satisfaction of energy conditions}. The article is then organized as follows: In Sec.~(\ref{Sol}) we briefly review the field equations of EGB gravity and consider the general static spherically symmetric metric for the wormhole spacetime. In Sec.~(\ref{sragy}) we seek to find wormhole solutions with Casimir source. Section (\ref{two}) is devoted to energy conditions and in Sec.~(\ref{Eqcon}) we deal with stability of the obtained solutions. Finally, in Sec.~(\ref{parti}) we study trajectory of particles in the wormhole spacetime and our conclusions are drawn in Sec.~(\ref{conclu}).
\section{Action and Field Equations}\label{Sol}
Firstly we give a brief introduction to the action of gravity in the framework of Lovelock theory. Up to the second order Lovelock term the action is given as 
\begin{equation}
	I_{g}=\int d^{n}x\sqrt{-g}\biggl[R-2\Lambda +\alpha _{2}\mathcal{L}_{GB}\biggr],\label{action}
\end{equation}
which is used in the low frequency limit of string theory and is added as the correction term to GR~\cite{myg1}. In the above action $n$
is the dimension of the spacetime, $\alpha_{2}$ being the Gauss-Bonnet (GB) coefficient, $R$ is the Einstein term and $\mathcal{L}_{GB}$ is GB Lagrangian
density which has the form%
\be
\mathcal{L}_{GB}=R^{2}-4R_{\mu \nu }R^{\mu \nu }+R_{\mu \nu \rho \sigma
}R^{\mu \nu \rho \sigma }.
\ee
In Lovelock theory, for each Euler density of order $k$ in $n$ dimensional spacetime, only the terms with $k<n$ exist within equations of
motion~\cite{shek}. Therefore, the solutions of the {EGB} theory are in $n\geq5$ dimensions. Varying the above action with respect to metric one obtains the field equation as
\begin{equation}\label{egbfield}
	G_{\mu \nu }^{(E)}+\alpha _{2}G_{\mu\nu}^{(GB)}=T_{\mu \nu },
\end{equation}%
where $T_{\mu \nu }$ is energy-momentum tensor (EMT), $G_{\mu \nu }^{(E)}$ is the Einstein tensor and $G_{\mu \nu }^{(GB)}$ is GB tensor which is of the form
\begin{equation}\label{egbten}
	G_{\mu \nu }^{(GB)}=2\left(-R_{\mu \sigma \kappa \tau }R_{\phantom{\kappa \tau
			\sigma}{\nu}}^{\kappa \tau \sigma }-2R_{\mu \rho \nu \sigma }R^{\rho \sigma
	}-2R_{\mu \sigma }R_{\phantom{\sigma}\nu }^{\sigma }+RR_{\mu \nu }\right)-\frac{1}{%
		2}\mathcal{L}_{GB}g_{\mu \nu }.
\end{equation}
In this paper, we are interested in investigating traversable wormholes which are introduced for the first time by Morris and Thorne~\cite{mt}. The line element for these {wormhole spacetimes} is given as%
\begin{equation}
	ds^{2}=-e^{2\phi (r)}dt^{2}+\left[ \frac{dr^{2}}{1-\frac{b(r)}{r}}%
	+r^{2}d\Omega _{n-2}^{2}\right],\label{metr1}
\end{equation}%
where $d\Omega _{n-2}^{2}$ is the metric on the surface of a $(n-2)$-sphere with  $d\Omega^2_1=d\varphi^2$ and $d\Omega^2_{i+1}=d\theta^2_i+\sin^2\theta_id\Omega^2_i$ for $i\ge1$ . Also, $\phi(r)$  and $b(r)$ are redshift and shape functions, respectively. The
conditions on $\phi (r)$ and $b(r)$ under which wormholes are traversable were discussed completely in~\cite{mt}. The redshift function should
be finite everywhere so that there are no singularity and event horizon in the spacetime. The shape function should satisfy the flaring-out condition i.e. $rb^{\prime}-b<0$ and $b(r)-r\leq 0$ (note that equality occurs only at throat of the wormhole denoted by $r_{0}$). The {EMT} in an orthogonal frame is given by
\be\label{emtt}
{T}_{\nu }^{\mu }=\mathrm{diag}\left[-\rho\left(r\right),p_{r}\left(r\right),p_{t}\left(r\right),p_{t}\left(r\right),...\right],
\ee
where $\rho (r)$ is the energy density and $p_{r}(r)$ and $p_{t}(r)$ are the radial and transverse pressures, respectively. Using the field equation (\ref{egbfield}), the EMT components then read 
\bea
\rho(r)&=&\frac{(n-2)}{2r^{2}}\Bigg\{\left( 1+\frac{2\alpha b}{r^{3}}\right) \frac{(rb^{\prime }-b)}{r}+\frac{b}{r}\left[(n-3)+(n-5)\frac{\alpha b}{r^{3}}\right]\Bigg\},  \label{rho} \\
p_{r}(r)&=&\frac{(n-2)}{2r}\Bigg\{2\left( 1-\frac{b}{r}\right) \left( 1+\frac{2\alpha b}{r^{3}}\right)\phi^{\prime }-\frac{b}{r^{2}}\left[ (n-3)+(n-5)\frac{\alpha b}{r^{3}}\right]\Bigg\},\label{pr}
\eea
\bea
\!\!\!\!\!\!p_{t}(r)\!\!\!&=&\!\!\!\left(1-\frac{b}{r}\right)\left(1+\frac{2\alpha b}{r^{3}}\right)\left[\phi ^{\prime \prime }+{\phi^{\prime }}^{2}+\frac{
	(b-rb^{\prime})\phi^{\prime}}{2r(r-b)}\right]+\left(1-\frac{b}{r}\right)\left( \frac{\phi^{\prime}}{r}+\frac{b-b^{\prime}r}{2r^{2}(r-b)}\right)\left[ (n-3)+(n-5)\frac{2\alpha b}{r^{3}}\right]\nn&-&\frac{b}{2r^{3}}\left[ \left( n-3\right) \left( n-4\right) +\left(
n-5\right) \left( n-6\right) \frac{\alpha b}{r^{3}}\right] -\frac{2\phi^{\prime}\alpha }{r^{4}}\left( 1-\frac{b}{r}\right)
\left(b-b^{\prime}r\right)\left(n-5\right),\label{pt}
\eea
where a prime denotes a derivative with respect to the {radial} coordinate $r$. We use a unit system with $8\pi G=1$ and also define $\alpha =(n-4)(n-3)\alpha _{2}$. Also, for the line element (\ref{metr1}), the Kretschmann and Weyl scalars are given as
\begin{eqnarray}
	{\mathcal K}&=&\mathcal{R}_{\mu \nu \gamma \beta}\mathcal{R}^{\mu \nu \gamma \beta}=\f{(n-2)}{r^6}\left(r^2b^{\prime2}-2rbb^\prime+2{b}^{2}n-5{b}^{2} \right)-\f{4{\phi^{\prime}}^3}{r^3}\left(r-b\right)\left(rb^\prime-b\right)+\f{4{\phi^{\prime}}^4}{r^2}\left(r-b\right)^2\notag \\
	&+&\f{\phi^{\prime}\phi^{\prime \prime}}{r^3}\left[8\phi^{\prime} r \left(r-b\right)-4rb^\prime+4b\right]\left[r-b\right]+{\frac {4{\phi^{\prime\prime}}^2 \left( b-r \right) ^{2}}{{r}^{2}}}\notag \\
	&&+\f{\phi^{\prime2}}{r^4}\left[{r}^{2}b^{\prime2}-2rbb^\prime+ \left( 4n-7 \right) {b}^{2}-8r \left( n-2 \right) b+4{r}^{2} \left( n-2 \right)\right],\label{kr11}
\end{eqnarray}
\begin{align}
	{\mathcal C}={C}_{\mu \nu \gamma \beta}{C}^{\mu \nu \gamma \beta}=\f{(n-3)}{(n-1)r^6}\left[4\phi^{\prime\prime}{r}^{2}\left( r-b \right) +4\phi^{\prime2}r^{2}\left(r-b\right)-2r\left(b^\prime r-3b+2r \right)\phi^{\prime}-12b+2b^\prime r\right].\label{ric1}
\end{align}
\section{Casimir Wormhole Solutions}\label{sragy}
We now have a set of three differential equations, i.e., the field equations (\ref{rho})-(\ref{pt}), with the five unknown functions $\rho(r)$, $p_{r}(r)$, $p_{t}(r)$, $b(r)$ and $\phi(r)$. Therefore, in order to determine the wormhole geometry, one can adopt several strategies. For instance, one can consider a specific equation of state (EoS) relating the EMT components, such as specific equations of state responsible for the present accelerated expansion of the Universe~\cite{pae1} and the {EoS for trace-less EMT}~\cite{kar95}. Here, we may apply some restrictions on the EMT components, e.g., taking the energy density as a predetermined function of radial coordinate. In this section we investigate wormhole configurations for which the supporting energy density is sourced by the Casimir effect. The energy associated with this effect has been computed for numerous flat and curved boundary configurations~\cite{casieffect}. Research over recent decades has demonstrated that the form of the Casimir energy and the resulting attractive or repulsive force {depend} critically on several factors: $1)$ the geometry and separation of the objects in the Casimir setup, $2)$ the type of quantum field considered (e.g., electromagnetic, fermionic, or scalar), and $3)$ the boundary conditions imposed, such as Dirichlet or Neumann conditions for a scalar field~\cite{casieffect,DDDNbcs,DDDNbcs1}, conductor conditions for the electromagnetic field~\cite{DDDNbcs1}, or bag model conditions for a spinor field~\cite{milclass}.
{More precisely}, the Casimir energy density for a scalar field confined inside a cube is $\rho_{\rm c}=-0.015/a^4$, where $a$ is the side length~\cite{Mamaev1979}. For wedge-shaped geometries formed by two conducting planes intersecting at an angle $\theta$, the energy densities for the electromagnetic and scalar fields are respectively given by~\cite{DD7879}
\bea\label{wedgecas}
\rho_{\rm c}(r)=-\f{1}{720\pi^2r^4}\left(\f{\pi^2}{\theta^2}-1\right)\left(\f{\pi^2}{\theta^2}+11\right),~~~~~~~~~~~~\rho_{\rm c}(r)=-\f{1}{1440\pi^2r^4}\left(\f{\pi^4}{\theta^4}-1\right),
\eea
with $r$ measured from the intersection line. For a spherical shell of radius $R$, the electromagnetic Casimir energy density near the surface ($D=R-r \ll R$) behaves approximately as $\rho_{\rm c}(r)\approx {\rm const}/RD^3$~\cite{Ola81,Mos88}. Further geometries have also been studied, including scalar fields for two spheres or a sphere and a plane~\cite{Bulgac2006}, a cylindrical shell~\cite{Raad81,Kitson2006}, two parallel cylinders, and a cylinder parallel to a plane~\cite{Teo2011}; additional details can be found in~\cite{casieffect,DDDNbcs1,Mil2004}. A common feature emerging from these examples is that the Casimir energy density typically scales with inverse powers of a characteristic distance within the configuration. Motivated by the above considerations, we adopt a general form for the Casimir energy density as
\begin{align}
	\rho(r)=\frac{\lambda}{r^m},\label{rhob}
\end{align}
where $\lambda$ and $m>0$ are constant parameters which depend on the type of quantum field, shape of the objects engaged in a Casimir setup and boundary conditions. The $\lambda$ parameter may also assume positive or negative values for different combinations of Dirichlet (D) and Neumann (N) boundary conditions, e.g. in the case of two parallel cylinders outside each other one gets $\lambda<0$ for DD and NN boundary conditions and $\lambda>0$ For DN and ND boundary conditions~\cite{DDDNbcs,DDDNbcs1}. In the former case the corresponding Casimir force is attractive and in the latter it is repulsive. In the case of higher dimensions, authors of~\cite{wolf} have calculated the Casimir energy density between two
parallel plates and for the sphere-plate geometry the Casimir energy has been evaluated in~\cite{teo11}. For instance, in the case of infinitely large, neutral, parallel ideal-metal planes at zero temperature separated by a distance $r$ in $n$ dimensions, the Casimir energy density is proportional to $r^{-n}$, see~\cite{wolf} for further details. Now, substituting for $\rho(r)$ into Eq.~(\ref{rho}) one obtains the following differential equation
\begin{align}
	{r}^{m+1} \left[ {r}^{3}+2\,{\it \alpha}\,b \left( r \right)  \right]\left(n-2 \right) b^{\prime}(r)+b(r)\left[\alpha(n-7) b(r) +{r}^{3} \left(n-4 \right)\right] \left( n-2 \right) {r}^{m}-2\,\lambda{r}^{6}=0. \label{cons2}
\end{align}
Next, we proceed to build and study wormhole solutions through solving the above differential equation for different values of the parameter $m$. We then define new parameter $\eta=(n-2)(m-n+1)$ that classifies the solutions with more details. 
\subsection{Case $\eta \neq 0 $}
Equation~(\ref{cons2}) is a rational ordinary differential equation with symmetries and {one can construct an integrating factor of the form $r^{n-m-8}$ and subsequently employ the total derivative of the term $(n-2)\left[r^{n-4}b(r) + \alpha r^{n-7}b(r)^2\right]$ with respect to $r$, with $\eta \neq 0$, such that the shape function $b(r)$ can be obtained as follows.} 
\begin{align}
	\alpha\eta b(r)^{2} {r}^{m-6}+\eta b(r){r}^{m-3}+{r}^{-n+m+1}{C}_1+2\lambda=0,\label{constrai12}
\end{align}
where the integration constant $C_1$ can be determined from the boundary condition $b(r_0)=r_0$ at the throat. We therefore obtain $C_1$ with the condition $m\ne n-1$ as 
\begin{align}
	C_1=-2r_0^{n-m-1}\lambda-\eta\left(\alpha{r_0}^{n-5}+{r_0}^{n-3}\right).
\end{align}
By solving the quadratic equation (\ref{constrai12}), the general solution is given by
\begin{equation}
	b_\pm(r)=\frac{r^3}{2\alpha \eta}\left[-\eta \pm \sqrt{\eta\left[\eta-4\alpha\left(C_1 r^{1-n}+2\lambda r^{-m}\right)\right]}\right].\label{ein1}
\end{equation}
Now, using Eq.~(\ref{cons2}), we find the flaring-out condition at the throat as
\begin{align}
	b^{\prime}(r_0)={\frac {\left(2-n\right) \left[\left(n-4\right)r_0^2+{\alpha}\left( n-7 \right)  \right]r_0^m+2\lambda
			r_0^4}{r_0^m\left(r_0^2+2{\alpha}\right)\left( n-2 \right)}}. \label{flare1}
\end{align}
In order to study the asymptotic behavior of this solution, one can expand Eq.~(\ref{ein1}) at infinity as
\begin{equation}
	1-\frac{b_{\pm}(r)}{r}\backsimeq 1+\frac{\big(\eta \mp \lvert\eta\rvert\big)r^{2}}{2\alpha \eta}\pm \frac{C_1 r^{3-n}+2\lambda r^{2-m}}{\lvert\eta\rvert}+{\mathcal O}\left(\frac{1}{r^{2n-4}}+\frac{1}{r^{2m-2}}+\frac{1}{r^{m+n-3}}\right).  \label{limitinf}
\end{equation}%
It is obvious that these solutions are asymptotically flat if we choose suitable signs in the equation above, namely, $\eta>0$ or $\eta<0$ with suitable values of the parameter $m$. We denote these solutions $b_{+}$ and $b_{-}$, respectively. Also, it is seen that for a suitable value of the parameters $m$,$\alpha$ and $\lambda$ we may have de-Sitter (small wormhole) and asymptotically flat or {asymptotically non-flat} solutions. In the following, these cases are discussed in detail.
\par
\begin{figure}[tbp]
	\begin{center}
		\includegraphics[width=0.48\textwidth,height=0.3\textheight]{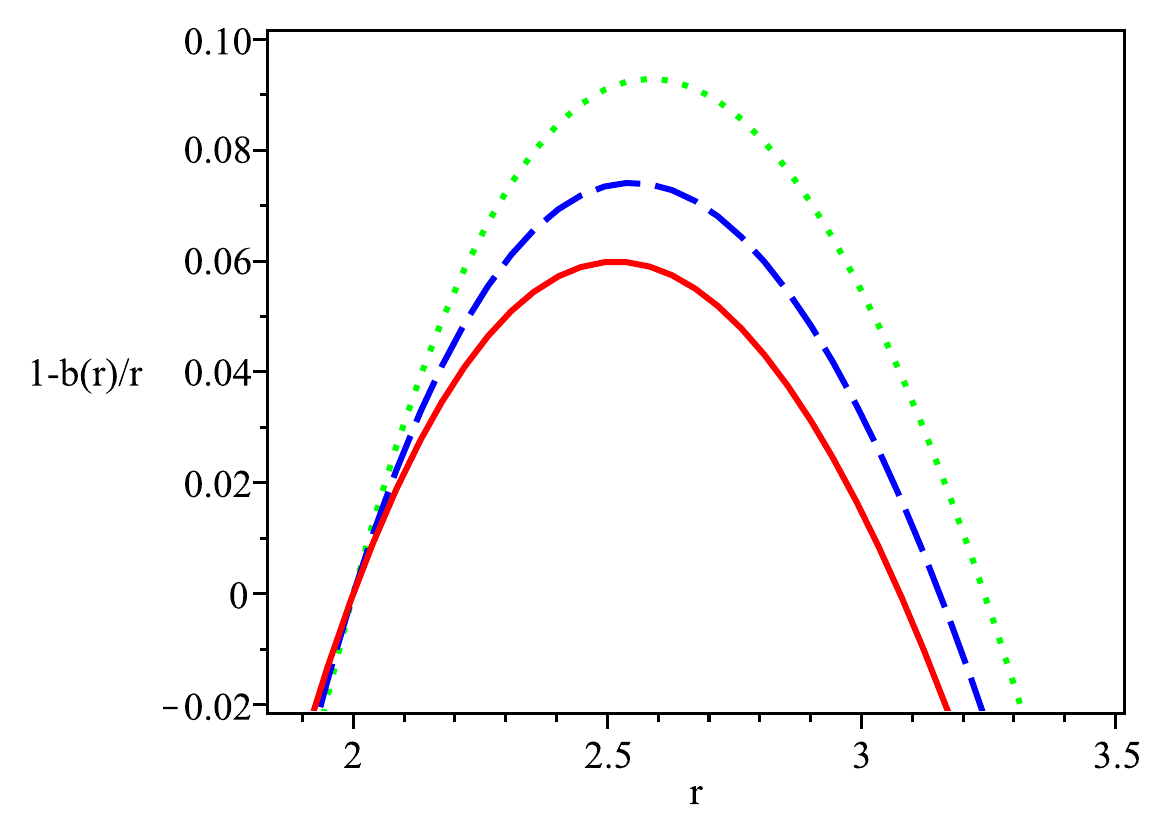}
		\includegraphics[width=0.48\textwidth,height=0.3\textheight]{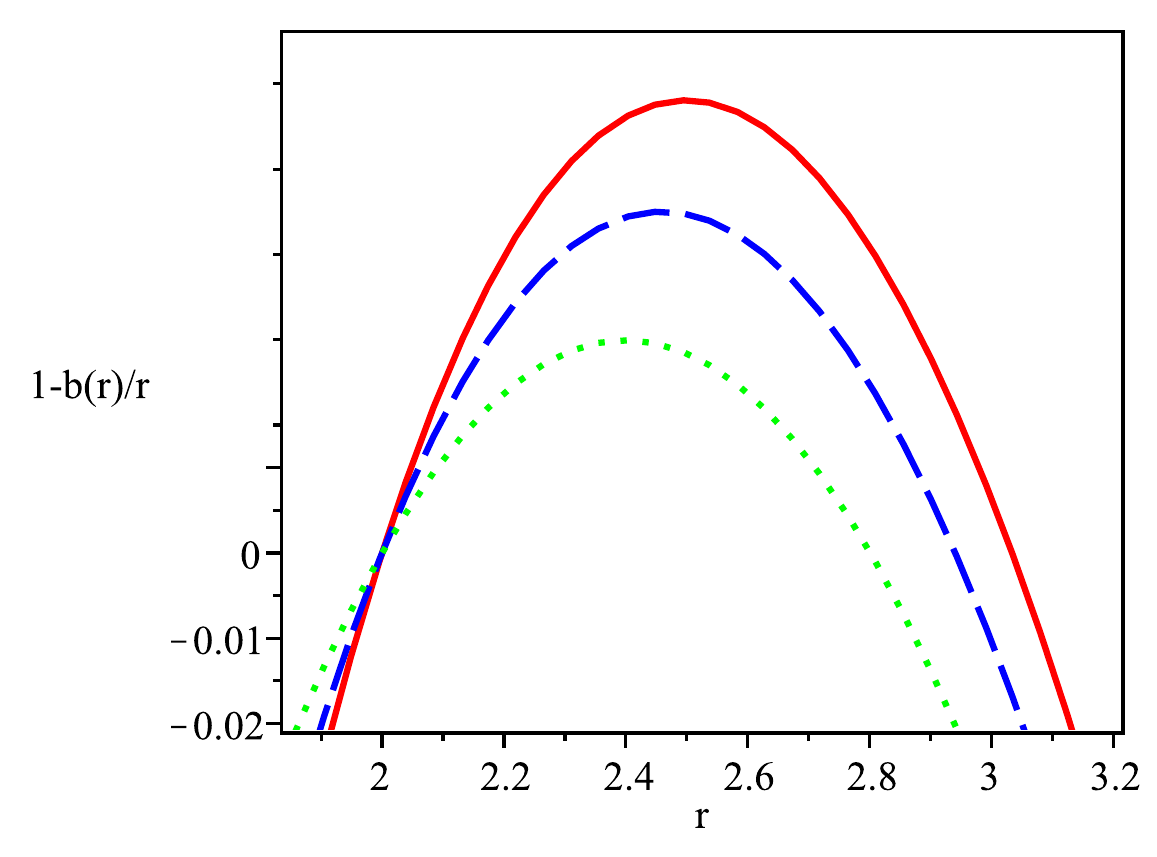}
	\end{center}
	\caption{The behavior of $1-b_{-}(r)/r$  versus $r$  for $\lambda=10$ , $\lambda=50$ , $\lambda=100$ from down to up  in the left panel and in right panel for $\lambda=-10$ , $\lambda=-50$ , $\lambda=-100$ from up to down. The constants are $m=7$, $\alpha=-13.3$ and $r_0=2$ in 7D.} \label{wor2}
\end{figure}

For the suitable values of $m$ and $\lambda$ parameters, one may obtain the wormhole solutions that cannot be arbitrarily large, or de Sitter-like wormhole solutions. we can choose the wormhole throat such that the condition $b^{\prime}(r_0)<1$ is satisfied. To be a solution of a wormhole, the condition $r-b(r)>0$ is also imposed. The condition $b(r_0)=r_0$ leads to two real and positive roots given by $r_{-}= r_0$ and $r_{+}$ that the latter satisfies the following equation
\begin{align}
	-\left[2r_0^{-m+n-1}\lambda+\left(\alpha r_0^{n-5}+r_0^{n-3}\right)\eta\right]r_{+}^7+2r_{+}^{-m+n+6}\lambda+\left(r_{+}^{4+n}+\alpha r_{+}^{2+n}\right) \eta=0.
\end{align}
Thus, the spatial extension of this type of wormhole configuration cannot be arbitrarily large and we have a finite wormhole within the range $r_- < r < r_+$. The left panel in Fig.~(\ref{wor2}) shows that an increase in the value of $\mid\lambda\mid$ parameter enlarges the wormhole
spatial extension. In contrast to this case, as we see in the right panel, a decrease in the value {of $\mid\lambda\mid$ parameter }reduces the wormhole
spatial extension.
\par
\begin{figure}
	\begin{center}
		\includegraphics[width=.48\textwidth]{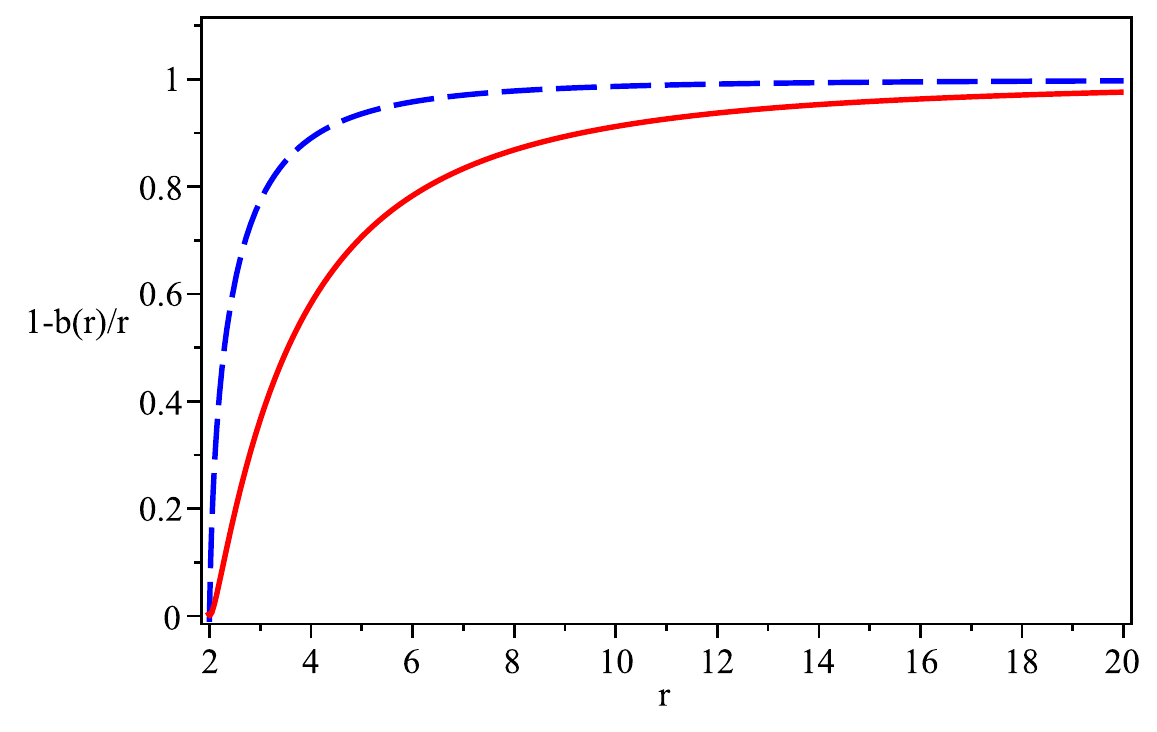}
		\includegraphics[width=.48\textwidth]{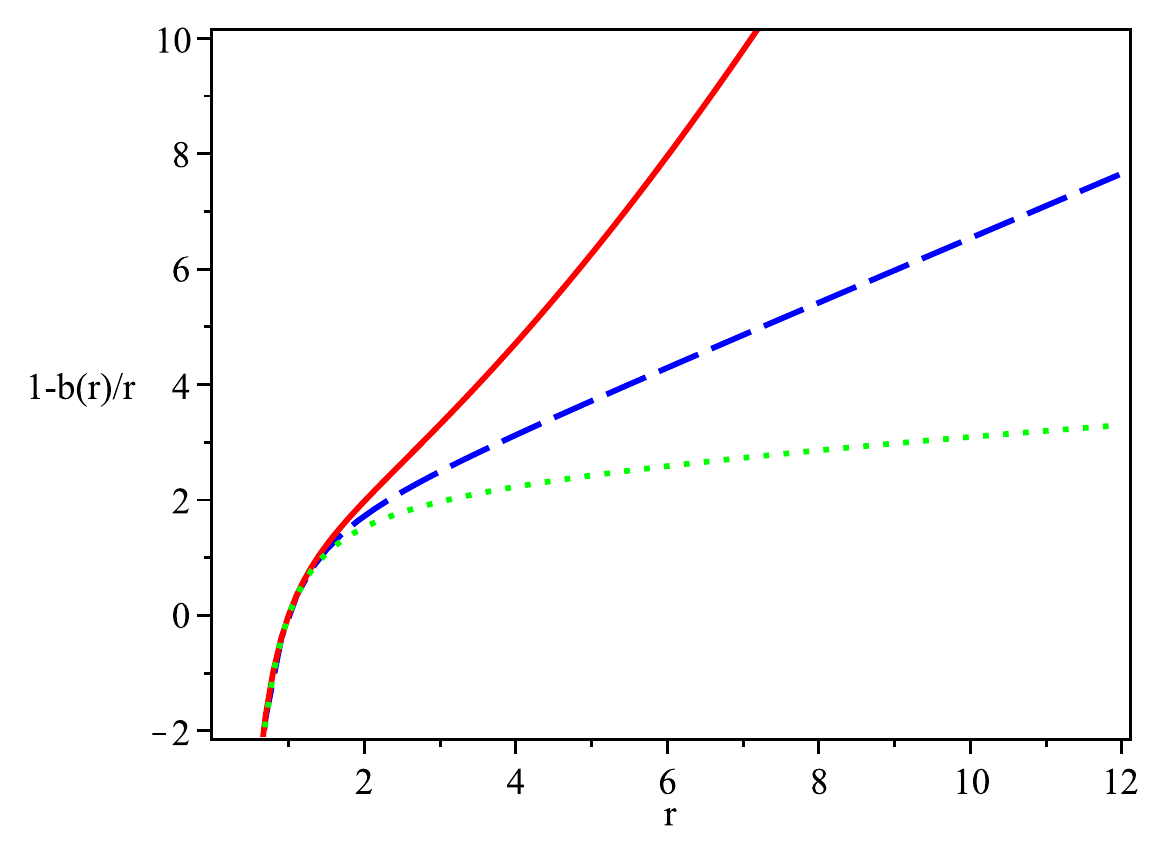}
		\caption{Left panel: The behavior of $1-\frac{b_{+}(r)}{r}$ versus $r$ for $\lambda =-4$ (dashed line), $\lambda=23.9$ (solid line) with $r_0=2$. We have set $\alpha=-1.6$ and $m=5$. Right panel: The behavior of $1-\frac{b_{-}(r)}{r}$ versus $r$ for $m =0.5$ (solid line), $m=1$ (dashed line) and $m=1.5$ (dotted line) with $r_0=1$. We have set $\alpha=0.002$ and $\lambda=-2.5$ in 5D.} \label{wor92}
	\end{center}
\end{figure}

For asymptotically flat wormhole solutions, i.e., those for which $b(r)/r\rightarrow0$ as $r$ tends infinity, we must choose the suitable signs in equation (\ref{limitinf}), namely, $\eta>0 $ for  positive solution ($b_{+}$), and $\eta<0 $ for negative solution $(b_{-})$. In addition, the condition $b(r_0)=r_0$ in Eq.~(\ref{constrai12}) for the real root $r_0$, determines the suitable values for the constant parameters $m$, $\alpha$ and $\lambda$. Imposing different values on $\lambda$, we can estimate the behavior of the inverse of radial metric component {$1-b_+(r)/r$ as shown in the left panel of Fig.~(\ref{wor92}), where the solution shows an asymptotically flat behavior}. The asymptotically non-flat solutions can be obtained by choosing $\eta<0$ in Eq.~(\ref{limitinf}) and hence, such type of solutions can appear in $b_-(r)$ branch. In such a situation, asymptotically we realize that $1-b_-(r)/r\propto r^{2-m}$. Therefore, asymptotically non-flat wormhole solutions can be obtained for $m<2$ and $\eta<0$. Note that as these solutions does not correspond to an asymptotically flat spacetime one may utilize junction conditions in order to match the interior wormhole solution to suitable exterior
vacuum spacetime~\cite{dmrub}. {The right panel of Fig.~(\ref{wor92}) shows the behavior of the quantity $1-b_{-}(r)/r$ for different values of the parameter $m$.}
\par
A fundamental challenge in GR is the prediction of spacetime singularities within its most physically meaningful solutions. A singularity represents a boundary or point in spacetime where quantities like density and the curvature of spacetime become infinite~\cite{singh}. This breakdown has profound consequences, calling into question the principles of physical determinism, causality, and the applicability of known laws of physics. The core issue is that GR, as a classical theory, ceases to be predictive in singular regions. When fundamental physical quantities diverge to infinity, the mathematical framework of the theory breaks down. This is widely interpreted not as a description of physical reality, but as a signal that GR is being applied beyond its limits of validity~\cite{JEr}. A spacetime singularity is commonly identified by its essential features, typically known as geodesic incompleteness (where the path of a particle or light ray cannot be extended beyond a certain point) and the unbounded growth of curvature invariants~\cite{essfeat,essfeat1,essfeat2}. This longstanding problem takes on a new and critical significance when considering wormhole structures connecting two separate Universes or distant regions of the same Universe. The presence of a singularity in a wormhole spacetime can pose the following problems or threats: $i)$ If a singularity exists in one of the Universes connected by the wormhole, the {\it infection} of broken physics could, in principle, travel through the wormhole to the other side. $ii)$ The most dangerous scenario is a timelike singularity, especially if it resides at the wormhole's throat. Contrary to a spacelike singularity (like the one inside a Schwarzschild black hole, which is unavoidable in the future~\cite{toolpoi}), a timelike singularity can be approached and then avoided. A hypothetical traveler could venture close to the singular throat, experience the breakdown of physics, and then escape back into one of the Universes, potentially carrying those disruptions with his/her. $iii)$ Such an event would fundamentally endanger the predictability and consistency of the laws of physics not just in the immediate vicinity of the wormhole, but throughout the connected Universes~\cite{KrSing,SPRsing}. Therefore, for a wormhole to be considered a physically viable and reasonable structure, a fundamental requirement must be met: the complete absence of spacetime singularities at the throat of wormhole. To check this one can compute the curvature scalars such as the Kretschmann and Weyl scalars and study their regular behavior at the wormhole throat~\cite{KrSing}. For the obtained solutions, these quantities can be calculated using the condition $b(r_0)=r_0$ in Eqs.~(\ref{kr11}) and (\ref{ric1})
\begin{eqnarray}
	\mathcal{R}_{\mu \nu \gamma \beta}\mathcal{R}^{\mu \nu \gamma \beta}\Big|_{r_0}&=&\frac{\phi^{\prime}(r_0)\left[\left(  \left( n-3 \right)r_0^{2}+{\alpha}\left( n-5 \right)  \right)  \left( n-2 \right)r_0^{m}-2r_0^{4}\lambda \right]^{2}}{r_0^{2} \left(r_0^{2}+2{\alpha} \right) ^{2}\left( n-2 \right) ^{2}r_0^{2m}} \notag \\ && \frac{\left( n-1 \right)  \left( n-2 \right) ^{2} \left(  \left( n-3
		\right)r_0^{4}+2{\alpha}\left( n-3 \right)r_0^{2}+{\alpha}^{2} \left( n-1 \right)  \right)}{\left( n-2 \right)r_0^{4} \left(r_0^{2}+2{\alpha} \right) ^{2}} \notag \\ && \frac{-4\left(  \left( n-3 \right)r_0^{2}+{\alpha}\left(n-5 \right)  \right)r_0^{4} \left( n-2 \right) \lambda r_0^{m}+4r_0^{8}{\lambda}^{2}}{\left( n-2 \right)r_0^4\left( r_0^{2}+2\,{\alpha} \right) ^{2}r_0^{2m}},\end{eqnarray} 
and
\begin{eqnarray}
	{C}_{\mu \nu \gamma \beta}{C}^{\mu \nu \gamma \beta}\Big|_{r_0}&=&\frac{\Sigma_4^2}{\left( n-1 \right)r_0^4 \left(r_0^2+2{\alpha} \right) ^{2} \left( n-2 \right)^{2}r_0^{2m}},
\end{eqnarray}
where
\begin{eqnarray}
	\Sigma_4&=& \left( n-3 \right) {r_0}\, \left[  \left(  \left( n-3 \right) r_0^{2}+{\alpha}\left( n-5 \right)  \right)  \left( n-2\right) r_0^{m}-2\lambda r_0^{4}\right]\phi^{\prime}\left(r_0\right) \notag \\ &&- \left( n-3 \right)  \left[ \left( n-1 \right)  \left( n-2 \right) 
	\left(r_0^{2}+{\it \alpha} \right)r_0^m+2\lambda r_0^{4} \right].
\end{eqnarray}
We therefore note that both of the above scalar invariants exhibit finite values at the throat when $r_{0}^2 \ne -2\alpha$ {and thus the spacetime singularity is absent at the throat.}
\subsection{Case $\eta=0$}
\begin{figure}[h]
	\begin{center}
		\includegraphics[width=0.48\textwidth,height=0.3\textheight]{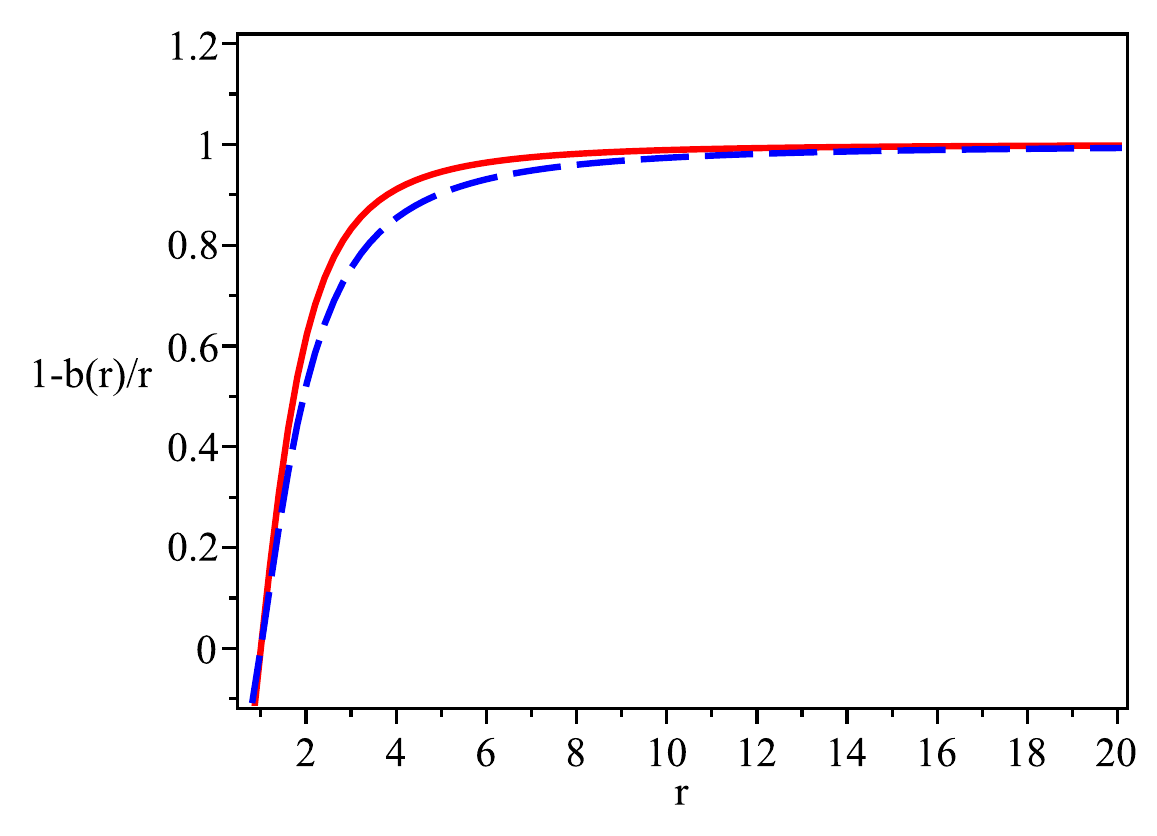}
		\includegraphics[width=0.48\textwidth,height=0.3\textheight]{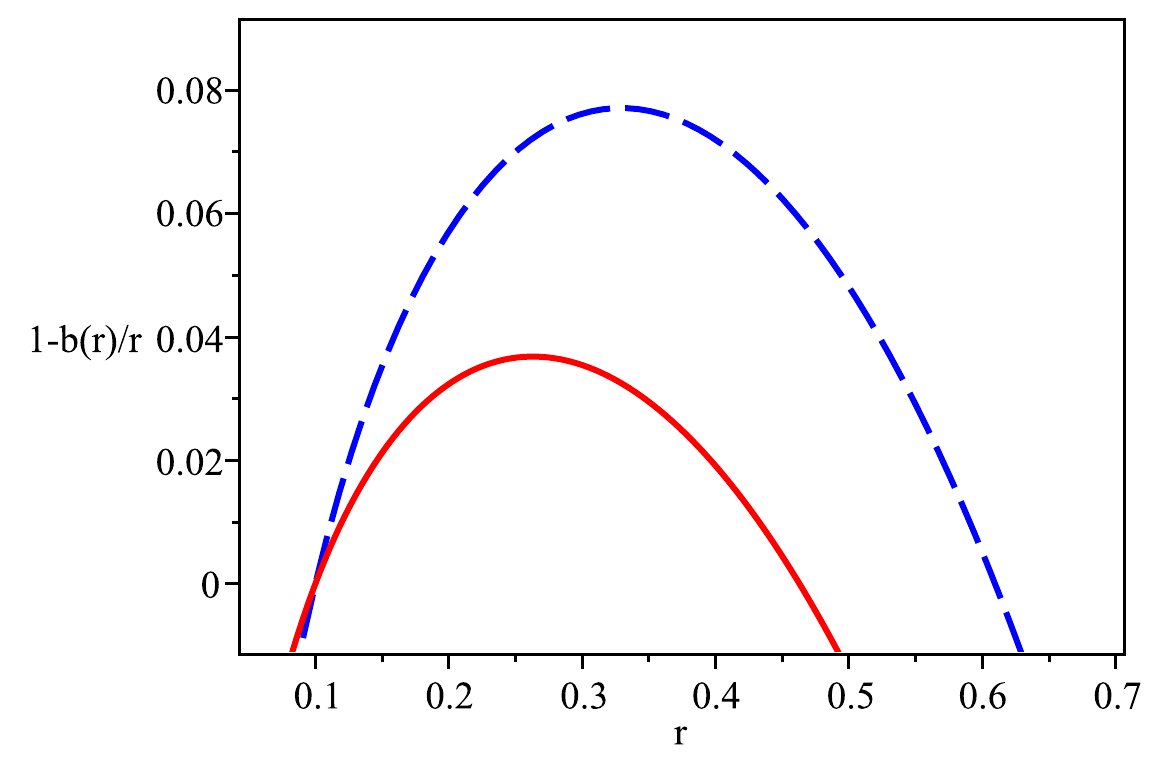}
	\end{center}
	\caption{Left panel: The behavior of $1-b_+(r)/r$ versus $r$ for $\lambda=0.5$, $\lambda=-0.5$ from down to up. The model parameters have been set as $m=4$, $\alpha=0.9$ and $r_0=1$ in 5D. Right panel: The behavior of $1-b_-(r)/r$ for $\lambda=0.3$ and $\lambda=0.2$ from up to down. The model parameters have been set as $m=4$, $\alpha=-1$ and $r_0=0.1$ in 5D.}\label{worm77}
\end{figure}

In the present subsection, we obtain wormhole solutions by considering the condition $\eta=0$ within the differential equation (\ref{cons2}), {and build an integrating factor as $r^{-7}$}  the solution is then found as
\begin{align}
	{\alpha}{r}^{n-7}(n-2) b \left( r \right)^{2}+ b(r)(n-2) {r}^{n-4}-2\lambda \ln(r)+C_2=0,
\end{align}
where the integration constant $C_2$ can be determined from the boundary condition $b(r_0)=r_0$ at the throat as
\begin{align}
	C_2=2\,\lambda\,\ln  \left(r_0\right)-\left(r_0^2+\alpha\right)\left(n-2\right)r_0^n-5.
\end{align}
By solving the quadratic equation, the general solution for the shape function is given by
\begin{equation}
	b_\pm(r)=\frac{[-1 \pm \sqrt{1-4\tilde{\alpha_2} r^{1-n}(C_2-2\lambda \ln(r))}]r^3}{2\alpha}, \label{ein3}
\end{equation}
where $\tilde{\alpha_2}=\frac{\alpha}{n-2}$. In order to study the behavior of this solution at infinity one can consider the following approximation
\begin{equation}
	1-\frac{b_{\pm}(r)}{r}\backsimeq 1+\frac{(1\mp 1) r^2}{2\alpha} \pm \frac{C_2-2\lambda \ln(r)}{(n-2)r^{n-3}}+{\cal O}\left(%
	\frac{1}{r^{2n-4}}\right). \label{limitinf2}
\end{equation}%
It is obvious that solutions with $b_{+}(r)$ are asymptotically flat and those with $b_{-}(r)$ can be dS small wormhole solutions with negative $\alpha$. For these solutions, we find the flaring-out condition by substituting $m=n-1$ in equation (\ref{flare1}). For the range of values $\lambda$ and $m$ the condition $b(r_0) = r_0$ leads to two real and positive roots given by $r_{-} = r_0$ and $r_{+}$ 
\begin{align}
	-{{r_0}}^{n-5} \left(r_0^{2}+{\alpha}\right)\left(n-2\right) +r_{+} \left(r_{+}^{2}+{\alpha} \right)\left( n-2 \right) 
	r_{+}^{n-6}+2\lambda\ln(r_0/r_+)=0,
\end{align}
and thus the spatial extension of this type of wormhole solution cannot be arbitrarily large. We plot the quantity $1-b_{+}(r)/r$ in the left panel of Fig.~(\ref{worm77}). Also, the right panel shows de-Sitter solutions for $1-b_{-}(r)/r$ with suitable choice of the model parameters. As the figure shows, an increase in the value $\mid\lambda\mid$ enlarges the wormhole spatial extension.  
\section{Energy Conditions}\label{two}
It is well-known that in GR framework, static traversable wormholes in four-dimensions violate the energy conditions at wormhole throat~\cite{hoc}. The flaring-out condition is sufficient for this result at or near the throat of the wormhole. {On the other hand, it is shown that the energy conditions can be satisfied in the vicinity of static wormhole throat in alternative gravity theories whose higher order terms of curvature are present~\cite{bhal,salhybpala}; this issue has been also addressed in the case of dynamic wormholes~\cite{meh}.} Among local energy conditions, we examined the weak energy condition (WEC) which requires that $T_{\mu \nu }v^{\mu }v^{\nu }\geq 0$ for the timelike vector field $v^{\mu}$. For a diagonal EMT, the WEC implies the following inequalities
\begin{equation}
	\rho \geq 0,~~~~~~\rho +p_{r}\geq 0,~~~~~~~~~\rho +p_{t}\geq 0,\label{E11}
\end{equation}%
to hold simultaneously. Note that the last two inequalities are defined as the null energy condition (NEC). In addition, the
strong energy condition (SEC) is satisfied through the following inequalities
\begin{equation}\label{E11}
	\rho+p_{r}+(n-2)p_{t} \geq0,~~~~~~~~~~~\rho+p_{r}\geq0,~~~~~~~~~~~\rho+p_{t}\geq0.
\end{equation}
Using now Eqs.~(\ref{rho})-(\ref{pt}) one obtains
\bea
\rho +p_{r}&=&-\f{(n-2)}{2r^{2}}\left[\f{b-rb^{\prime}}{r}+2\phi ^{\prime }\left( b-r\right) \right] \left(1+\frac{2\alpha b}{r^{3}}\right), \label{wec1} \\
\rho +p_{t}&=& -\frac{\left( b-rb^{\prime }\right) }{2r^{3}}\left( 1+\f{6\alpha b}{r^{3}}\right) +\f{b}{r^{3}}\left[ (n-3)+(n-5)\f{2\alpha b}{r^{3}}\right]   \nn
&+&\phi ^{\prime }\bigg[\f{b-rb^{\prime }}{2r^{2}}\left( 1-\frac{2\alpha b
}{r^{3}}(9-2n)\right) -\f{b}{r^{2}}\left( (n-3)+\frac{2\alpha b}{r^{3}}
(n-5)\right) \nn
&+& \f{1}{r}\bigg((n-3)+\f{2\alpha b^{\prime }(n-5)}{r^{2}}\bigg)\bigg]
+\left( 1-\frac{b}{r}\right) \left( 1+\frac{2\alpha b}{r^{3}}\right) \left( {
\phi ^{\prime }}^{2}+\phi ^{\prime \prime }\right),\\
\rho+p_{r}+(n-2)p_{t}&=&(n-2)\left( 1-\f{b}{r}\right) \left( 1+\f{2\alpha b}{r^{3}}\right) \left( {\phi ^{\prime }}^{2}+\phi ^{\prime \prime }\right)\nn
&-&\f{(n-2)\phi ^{\prime }}{r} \left[\alpha b^{\prime}\big(\f{2}{r^2}-\frac{3 b}{r^3}\big)+\f{(5-2n)}{r}+\frac{2\alpha(2n-10)}{r^3}-\frac{b^{\prime}}{2}+\f{\alpha(11-2n) b^2}{r^4}\right]\nn 
&-&\f{(n-2)(n-4)}{2 r^2}\left[b^{\prime}+\f{(n-4)b}{r}\right]-\f{(n-2)(n-6) \alpha b}{r^5}\left[ b^{\prime}+\frac{(n-7)b}{2r}\right].
\eea
One then can easily show that for $\alpha =0$ and $\phi ^{\prime }=0$, WEC is violated i.e., $\rho +p_{r}<0$ due to the flaring-out condition. Note that at the throat, one can verify that
\begin{eqnarray}
	\rho + p_r\Big|_{r=r_0}=-\frac{n-2}{2r_0^2}\left(1-b^{\prime }_0 \right)\left( 1+\frac{2\alpha}{r_0^2} \right) \,.
\end{eqnarray}
\begin{eqnarray}
	\rho + p_t\Big|_{r=r_0}=\frac{\phi_0^{\prime}}{2r_0}\left(1-b^{\prime }_0 \right)\left( 1+\frac{2\alpha}{r_0^2} \right) -\frac{1}{2r_0^2}\left(
	1-b^{\prime }_0 \right)\left( 1+\frac{6\alpha}{r_0^2} \right)+\frac{1}{{r_0}^2}\left[(n-3) +\frac{2\alpha(n-5)}{r_0^2} \right],
\end{eqnarray}
and
\begin{eqnarray}
	\rho +p_r+(n-2)p_t\Big|_{r=r_0}&=&\frac{(n-2)\phi_0^{\prime}}{2r_0}\left(
	1-b^{\prime }_0 \right)\left( 1+\frac{2\alpha}{r_0^2} \right) \notag \\
	&&+\frac{(n-2)}{2r_0^2}\left(
	1-b^{\prime }_0 \right)\left[ (n-4)+\frac{2\alpha(n-6)}{r_0^2} \right] \notag \\
	&&-\frac{(n-2)}{2{r_0}^2}\left[(n-3)(n-4) +\frac{2\alpha(n-5)(n-6)}{r_0^2} \right].
\end{eqnarray}
Taking into account the condition $b^{\prime }_0<1$, and for $\alpha>0$, one verifies the general condition $\left( \rho + p_r \right)\big|_{r=r_0}<0$. In order to impose $\left( \rho + p_r \right)\big|_{r=r_0}>0$, one needs to consider $\alpha<0$ and the condition $r_0<\sqrt{2\lvert\alpha\rvert}$, which proves that one may have wormholes in GB gravity satisfying the WEC and SEC with choosing $\phi^{\prime}<0$ at the throat.
The WEC profiles for energy density presented in {Eq.~(\ref{rhob})} are given by
\begin{eqnarray}
	\!\!\!\!\rho (r)+p_{r}(r)\!\!\!\!&=&\!\!\!\!\frac{\Sigma_5\phi^{\prime}- \left( n-2 \right) b \left[{\alpha}\left(n-5 \right) b +{r}^{3} \left( n-3 \right)\right] {r}^{m}+2\lambda{r}^{6}}{2 r^{6+m}},\\\label{wecc1}
	\!\!\!\!\rho (r)+p_{t}(r)\!\!\!\!&=&\!\!\!\!\frac{\Sigma_6\phi^{\prime}+\Sigma_7+[{r}^{m+2} \left( r-b \right)  \left[ {r}^{3}+2
		\alpha b \right] ^{2} \left( n-2 \right) 
		] \left(\phi^{\prime \prime}+{\phi^{\prime}}^2\right)}{\left( {r}^{3}+2\alpha b\right)  \left( n-2 \right)r^{6+m}},\\\label{wecc2}
	\!\!\!\!\rho (r)+p_{r}(r)+(n-2)p_{t}(r)\!\!\!\!&=&\!\!\!\!\frac{\Sigma_8+\Sigma_9\phi^{\prime}+[r \left( r-b \right) ^{2} \left( {r}^{3}+2\alpha b\right) ^{2} \left( n-2 \right) 
		]\left(\phi^{\prime \prime}+{\phi^{\prime}}^2\right)}{r^{5}\left( r-b\right)  \left( {r}^{3}+2\alpha b \right)},\label{wec3}
\end{eqnarray}
where
\begin{eqnarray}
	\Sigma_5&=&2\,{r}^{m+1} \left( r-b \right)  \left( {r}^{3}+2\alpha b  \right)  \left( n-2 \right),\\
	\Sigma_6 &=&  -\left[{\alpha}^{2} \left( n-5 \right)b^{3}+\f{r}{2}\left(  \left( n-9 \right) {r}^{2}+4\alpha\left( n-5 \right)  \right) \alpha b^2\right]  \left( n-2 \right) {r}^{m+1} \notag \\
	&&+\left[-\f{r^4}{2} \left(\left( n-3 \right) {r}^{2}-4\,\alpha\, \left( n-5 \right)  \right) b +{r}^{7}\left( n-3 \right)  \right]  \left( n-2 \right) {r}^{m+1}-\notag \\&& \left( 6\alpha b  +{r}^{2} \left( {r}^{2}-4\alpha\right)\right){r}^{6}\lambda,\\
	\Sigma_7&=&\f{1}{2}\left[2\alpha^{2} \left( n-5 \right)b^{2}+{r}^{3}\alpha\left( n-9\right) b + \left( n-3 \right) {r}^{6} \right] b\left( n-2 \right) {r}^{m}+{r}^{6}\lambda\left({r}^{3}+6 \alpha b\right),\\
	\Sigma_8&=&-2\, \left( n-5 \right) {r}^{6-m}{\alpha}\lambda b -{r}^{2}{\alpha}b(r)^2 \left( n-1 \right)  \left( n-2 \right)  \left( r-b\right) \notag \\
	&&+{r}^{8-m}\lambda \left(  \left( n-4 \right) b +r \right) -\lambda{r}^{9-m} \left( n-3 \right) +2\left(  \left( n-6 \right) b +r \right) {r}^{5-m}{\alpha}\lambda b,\\
	\Sigma_9&=&-2 \left( 2r-3b \left( r \right)  \right) {r}^{5}{\alpha}\lambda b {r}^{1-m}-{r}^{5-m}\lambda\,{r}^{5}+\lambda\,{r}^{5}b{r}^{4-m}+4{r}^{5}{\alpha}\,
	\lambda \left(r-\f{3b}{2}\right) {r}^{2-m}\notag \\
	&&-\Big[\alpha^{2} \left( n-1 \right) b^{3}+\f{r}{2}{\alpha}\left((n-1) {r}^{2}+
	4\,{\alpha}\left( n-3 \right)  \right)b^{2}\notag \\
	&&+\f{r^4}{2}\left(  \left( n-1 \right) {r}^{2}+4\,\alpha\, \left( n-3 \right)  \right) b-{r}^{7} \left( n-2 \right)\Big]  \left( n-2 \right)\left( r-b\right).
\end{eqnarray}
Let us now obtain the asymptotic behavior of the energy conditions, hence the expressions $\rho (r)+p_{r}(r)$ and $\rho (r)+p_{t}(r)$ in limit of large $r$ read 
\begin{equation}
	\rho (r)+p_{r}(r)\backsimeq \frac{(n-2)\phi^{\prime}}{r}+{\cal O}\left(\frac{1}{r^{m}}\right),  \label{limitinwec1}
\end{equation}
\begin{equation}
	\rho (r)+p_{t}(r)\backsimeq \left(\phi^{\prime \prime}+{\phi^{\prime}}^2\right)+\frac{(n-3)\phi^{\prime}}{r}-\frac{\lambda \phi^{\prime}}{(n-2)r^{m-1}}+{\cal O}\left(\frac{1}{r^{m+1}}\right).  \label{limitinwec11}
\end{equation}
Also for the asymptotic behavior of the strong energy condition we get
\begin{equation}
	\rho (r)+p_{r}(r)+(n-2)p_{t}(r)\backsimeq \left(\phi^{\prime \prime}+{\phi^{\prime}}^2\right)(n-2)+\frac{(n-2)^2\phi^{\prime}}{r}-\frac{\lambda \phi^{\prime}}{r^{m-1}}+{\cal O}\left(\frac{1}{r^{m+1}}\right).  \label{limitinwec12}
\end{equation}
It is therefore seen that the asymptotic behavior of $\rho (r)+p_{r}(r)$ and $\rho (r)+p_{t}(r)$ as well as $\rho (r)+p_{r}(r)+(n-2)p_{t}(r)$ depend on the first and second derivatives of the red shift function at spatial infinity. We then may choose a suitable form for the redshift function in order that the energy conditions are satisfied throughout the entire spacetime.
\subsection{Zero tidal-force solutions}
\begin{figure}[h]
	\begin{center}
		\includegraphics[width=0.48\textwidth,height=0.3\textheight]{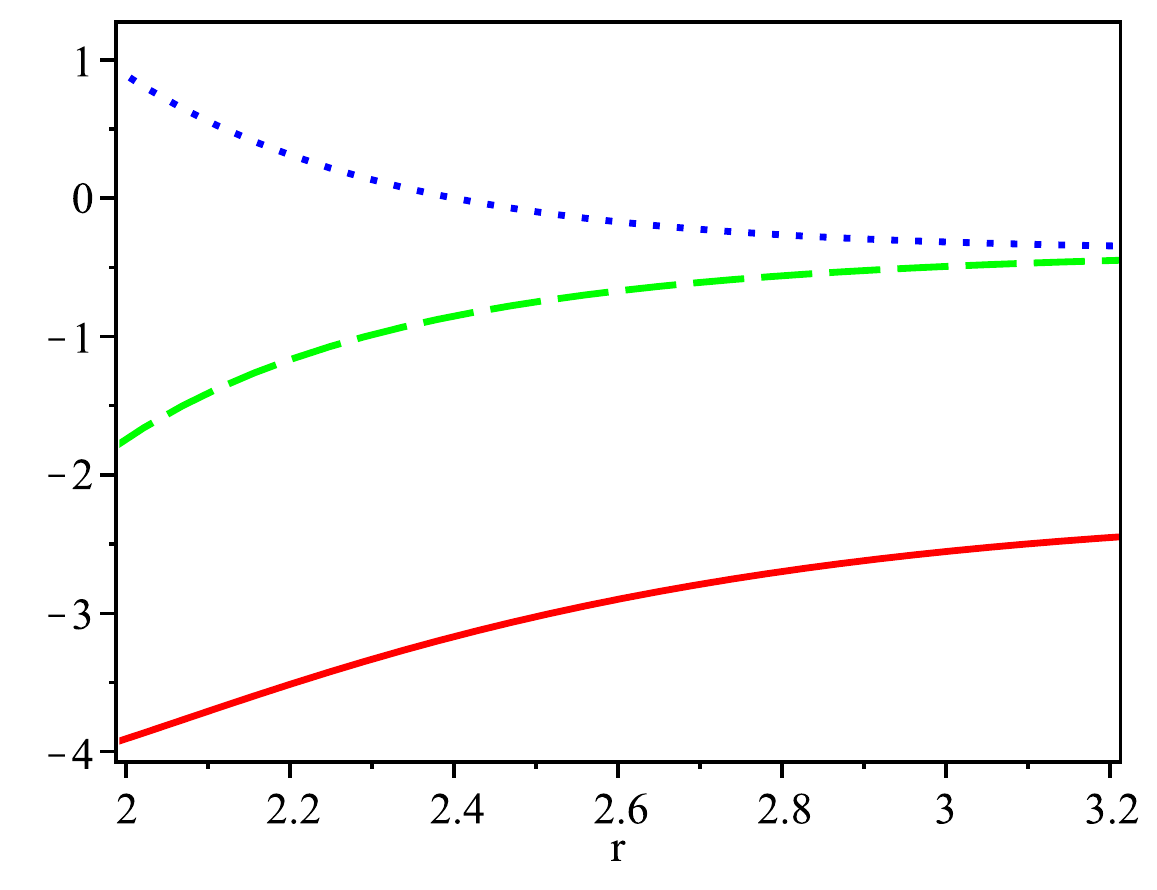}
		\includegraphics[width=0.48\textwidth,height=0.3\textheight]{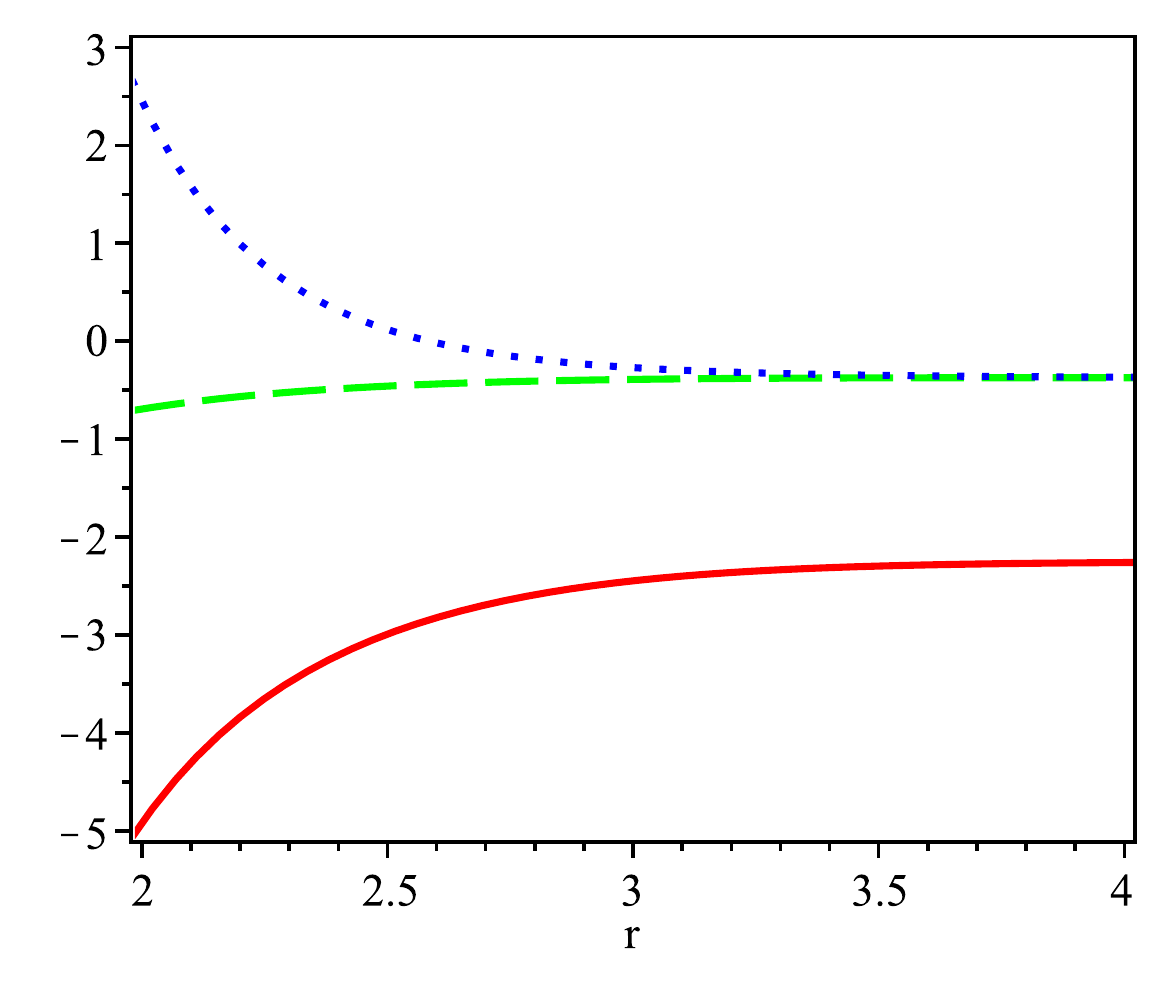}
	\end{center}
	\caption{The behavior of $\protect\rho  +p_{r}+5p_{t} $ (solid), $\protect\rho +p_{r}$ (dotted) and $\protect\rho +p_{t}$ (dashed) versus $r$ for $\lambda=-100$ (left panel) and $\lambda=100$ (right panel) respectively. The model parameters have been set as $\alpha=-13.3$, $m=7$, and $r_0=2$ in 7D.}
	\label{wor9}\label{wor6a}\label{worm7}
\end{figure}

\begin{figure}[h]
	\begin{center}
		\includegraphics[width=0.48\textwidth,height=0.3\textheight]{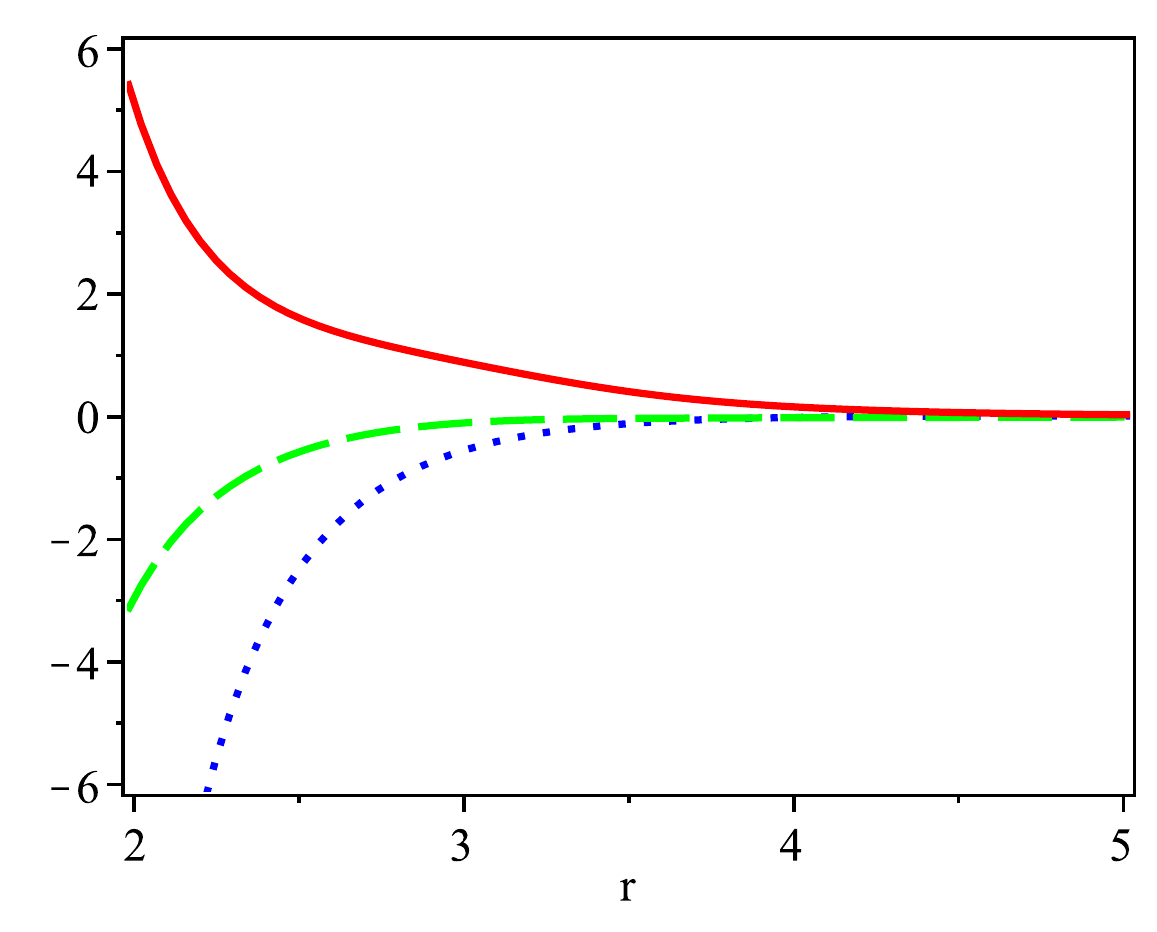}
		\includegraphics[width=0.48\textwidth,height=0.3\textheight]{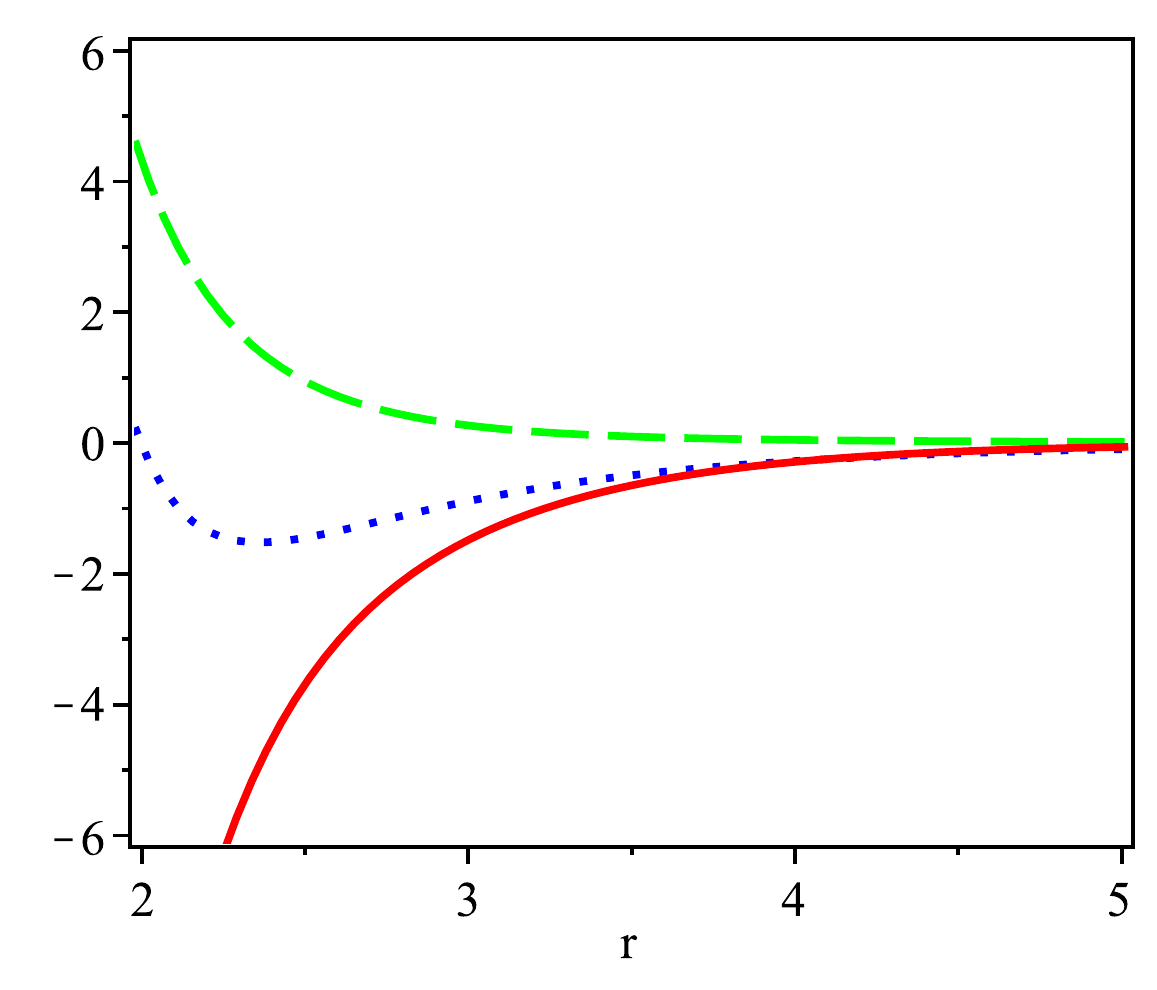}
	\end{center}
	\caption{The behavior of $\protect\rho  +p_{r}+5p_{t} $ (solid), $\protect\rho +p_{r}$ (dotted) and $\protect\rho +p_{t}$ (dashed) versus $r$ for $\lambda=-850$ (left panel) and $\lambda=850$ (right panel) respectively. The model parameters have been set as $\alpha=13.3$, $m=7$, and $r_0=2$ in 7D.}
	\label{wor9as}\label{wor6asa}\label{worm7as}
\end{figure}

\begin{figure}[h]
	\begin{center}
		{\includegraphics[width=.4\textwidth]{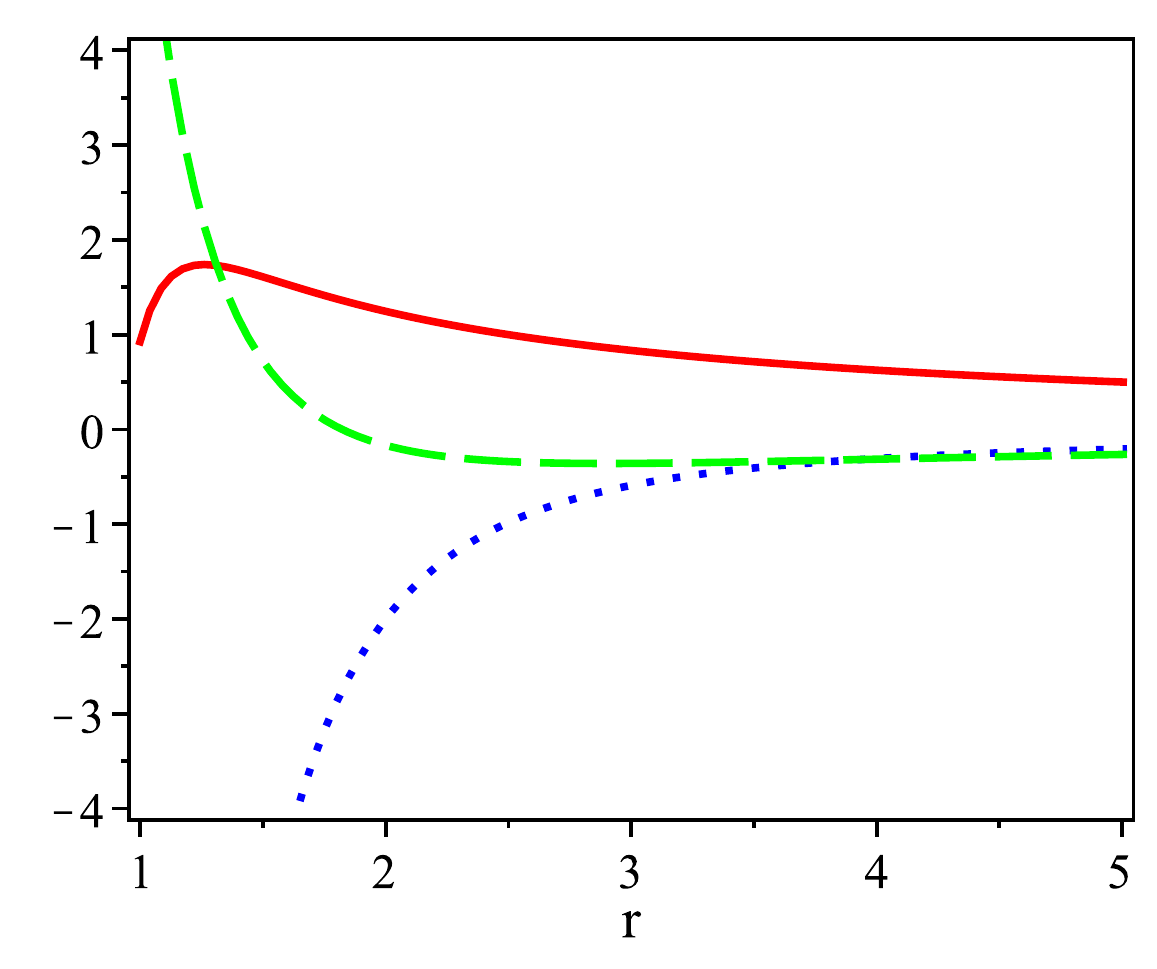}}
		\caption{The behavior of $\protect\rho  +p_{r}+5p_{t} $ (solid), $\protect\rho +p_{r}$ (dotted) and $\protect\rho +p_{t}$ (dashed) versus $r$ for  $\lambda=-2.5$. The model parameters have been set as $\alpha=0.002$, $m=1$, and $r_0=2$ in 5D.}
		\label{wecnonas1}
	\end{center}
\end{figure}

In section \ref{sragy}, properties of the shape functions were discussed in detail. In what follows, we discuss the energy conditions using Eqs.~(\ref{wecc1})-(\ref{wec3}) for these solutions. The first class of solutions deals with the case of constant redshift function, i.e., $\phi^{\prime}=0$. These solutions are called the zero-tidal-force solutions so that a stationary observer hovering about the gravitational field of the wormhole, will not experience any tidal force. For this case, the general solutions with $\eta \ne 0$ and $\eta=0$ were obtained and studied before. In Fig.~(\ref{wor9}), we show the behavior of the expressions $\rho (r)+p_{r}(r)$, $\rho (r)+p_{t}(r)$ and $\rho (r)+p_{r}(r)+5p_{t}(r)$  for the case of de-Sitter wormhole solutions. Note that all of the quantities $\rho (r)+p_{r}(r)+5p_{t}(r)$ and $\rho (r)+p_{t}(r)$ are negative throughout the spacetime. For asymptotically flat wormholes {as shown in the left panel of Fig. \ref{wor6asa}, assuming $\lambda<0$}, the quantities $\rho (r)+p_{t}(r)$ and $\rho (r)+p_{r}(r)$ are negative throughout the spacetime. As the right panel shows, the NEC is satisfied in the vicinity of the throat with a suitable choice of positive $\lambda$ parameter. In Fig.~\ref{wecnonas1} for asymptotically non-flat solutions the NEC and SEC are violated throughout the spacetime.


\subsection{Nonzero tidal-force solutions}
\begin{figure}[h]
	\begin{center}
		\includegraphics[width=0.48\textwidth,height=0.3\textheight]{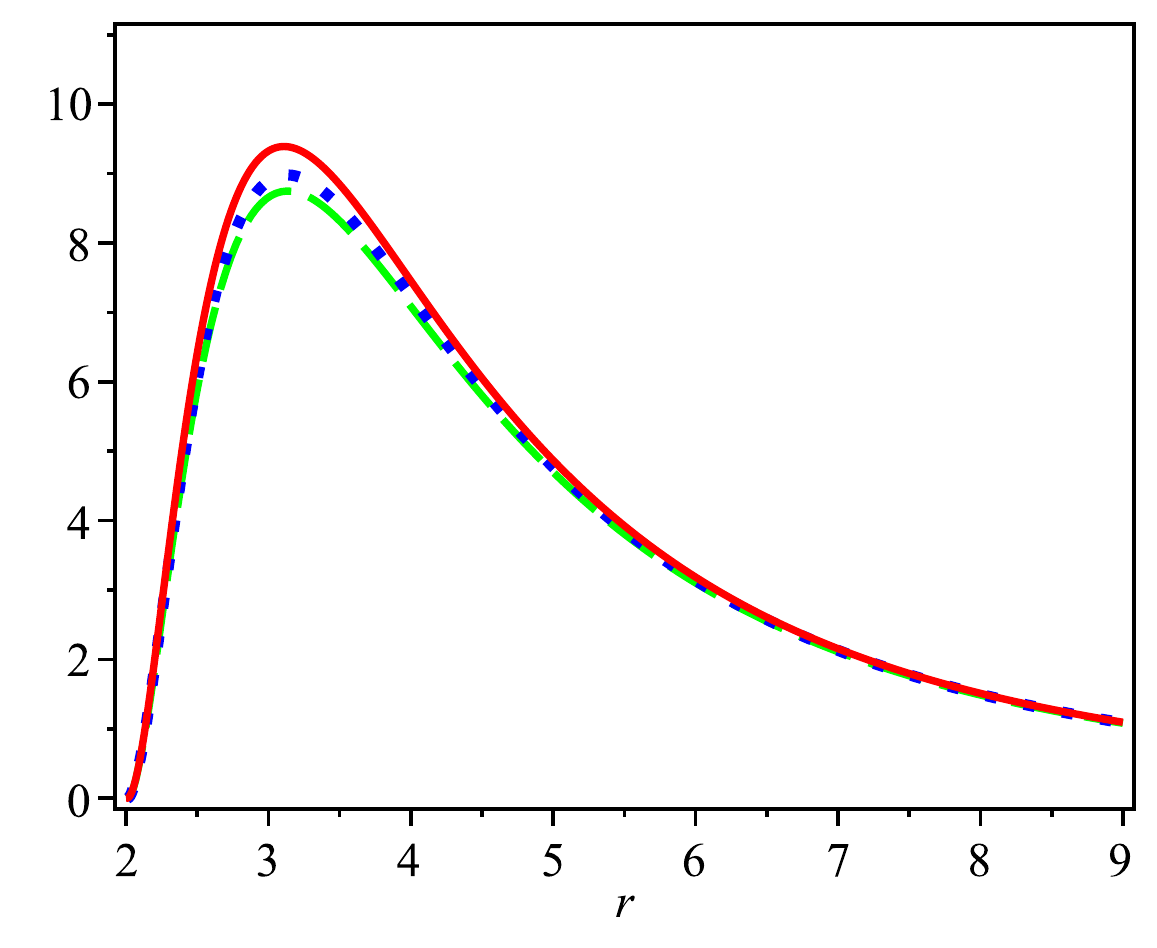}
		\includegraphics[width=0.48\textwidth,height=0.3\textheight]{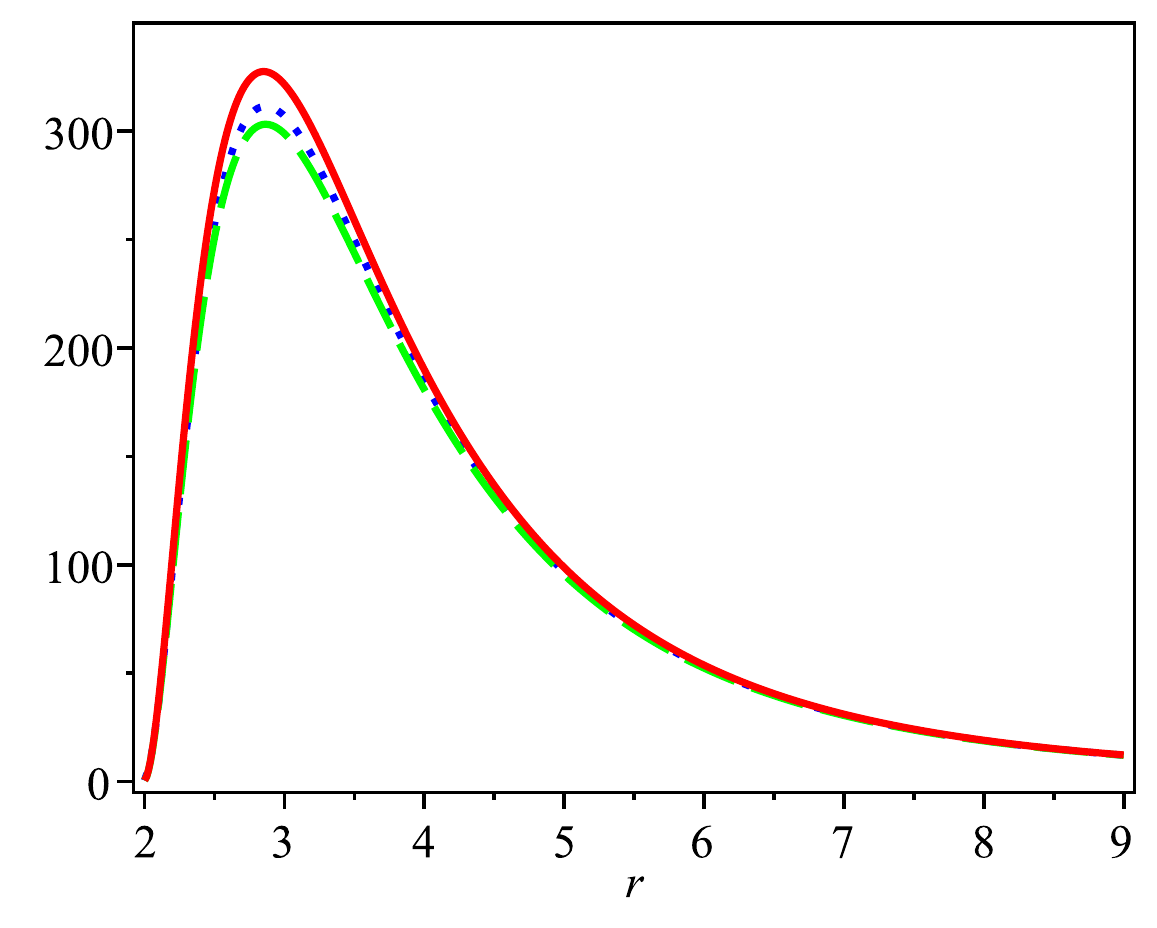}
	\end{center}
	\caption{The behavior of  $\protect\rho +p_{r}$  and $\protect\rho +p_{t}$ versus $r$ for  $\lambda=23.99$ and $k=1$ in the left and right panels, respectively. The model parameters have been set as $\alpha=-1.6$ (solid curve), $\alpha=-1$ (dotted curve), $\alpha=-0.5$ (dashed curve), $m=5$, and $r_0=2$ in 5D.}
	\label{wor9phi}\label{worm7phi}\label{wor6aphi}
\end{figure}

\begin{figure}[h]
	\begin{center}
		\includegraphics[width=0.48\textwidth,height=0.3\textheight]{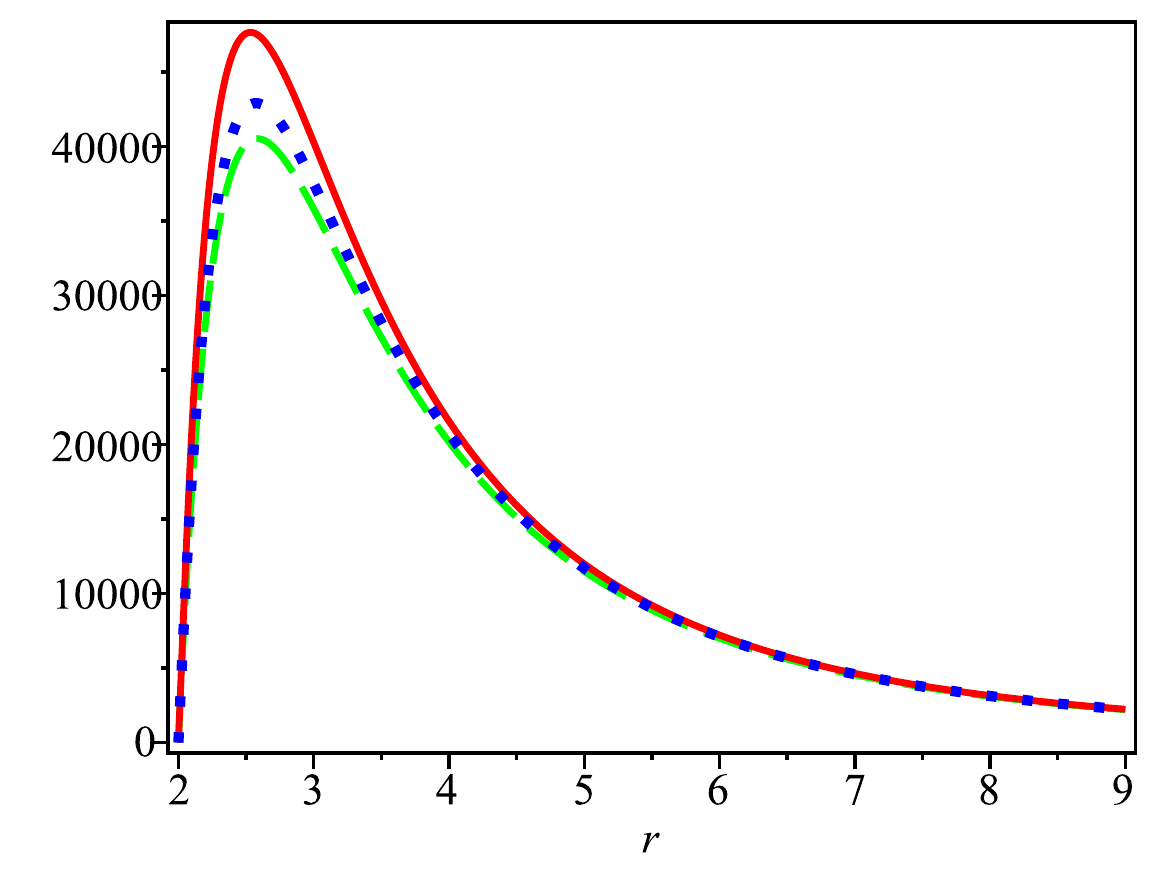}
		\includegraphics[width=0.48\textwidth,height=0.3\textheight]{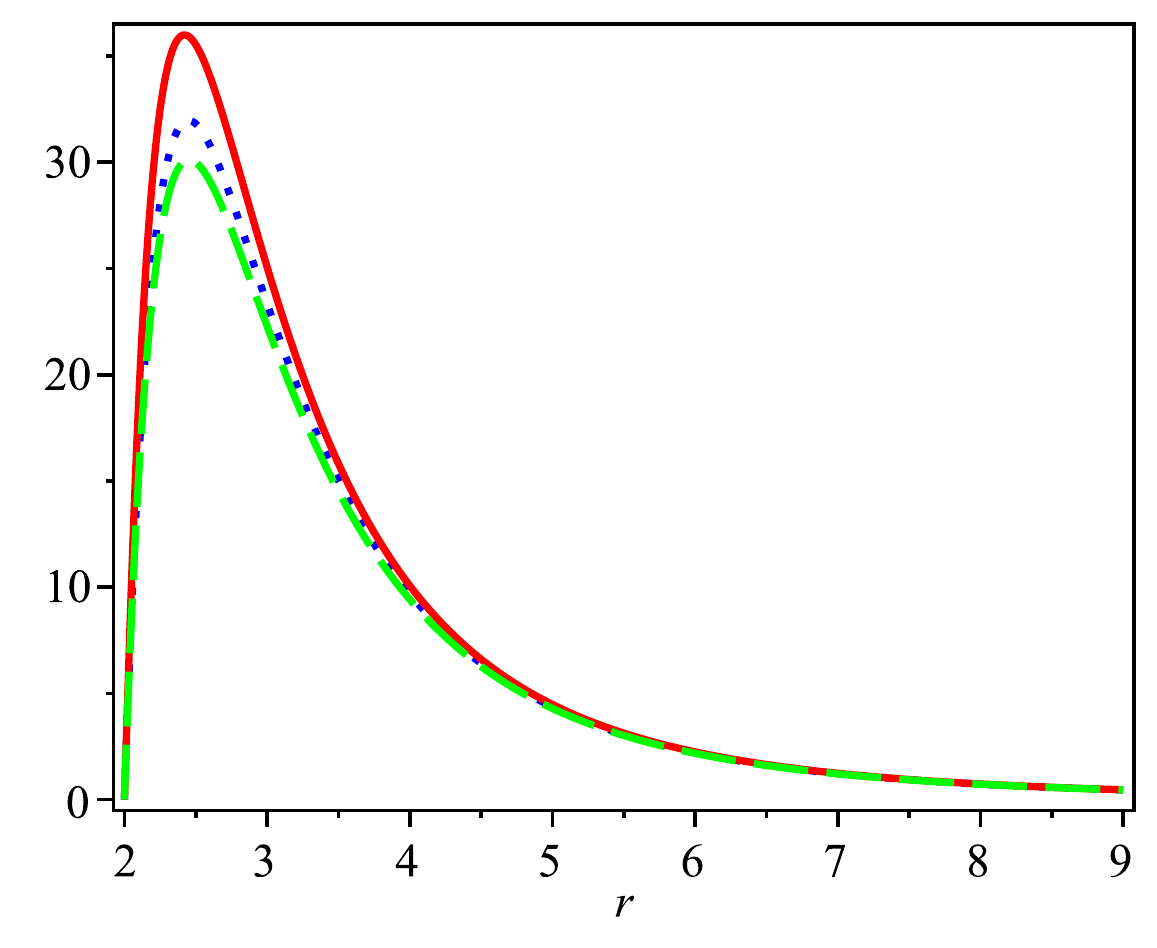}
	\end{center}
	\caption{The behavior of  $\protect\rho +p_{r}$  and $\protect\rho +p_{t}$ versus $r$ for  $\lambda=-0.0001$ and $k=1$ in the left and right panels, respectively. The model parameters have been set as $\alpha=-1.6$ (solid curve), $\alpha=-1$ (dotted curve), $\alpha=-0.5$ (dashed curve), $m=5$, and $r_0=2$ in 5D.}
	\label{wor9phinn}\label{worm7phinn}\label{wor6aphinn}
\end{figure}

\begin{figure}[h]
	\begin{center}
		\includegraphics[width=0.48\textwidth,height=0.3\textheight]{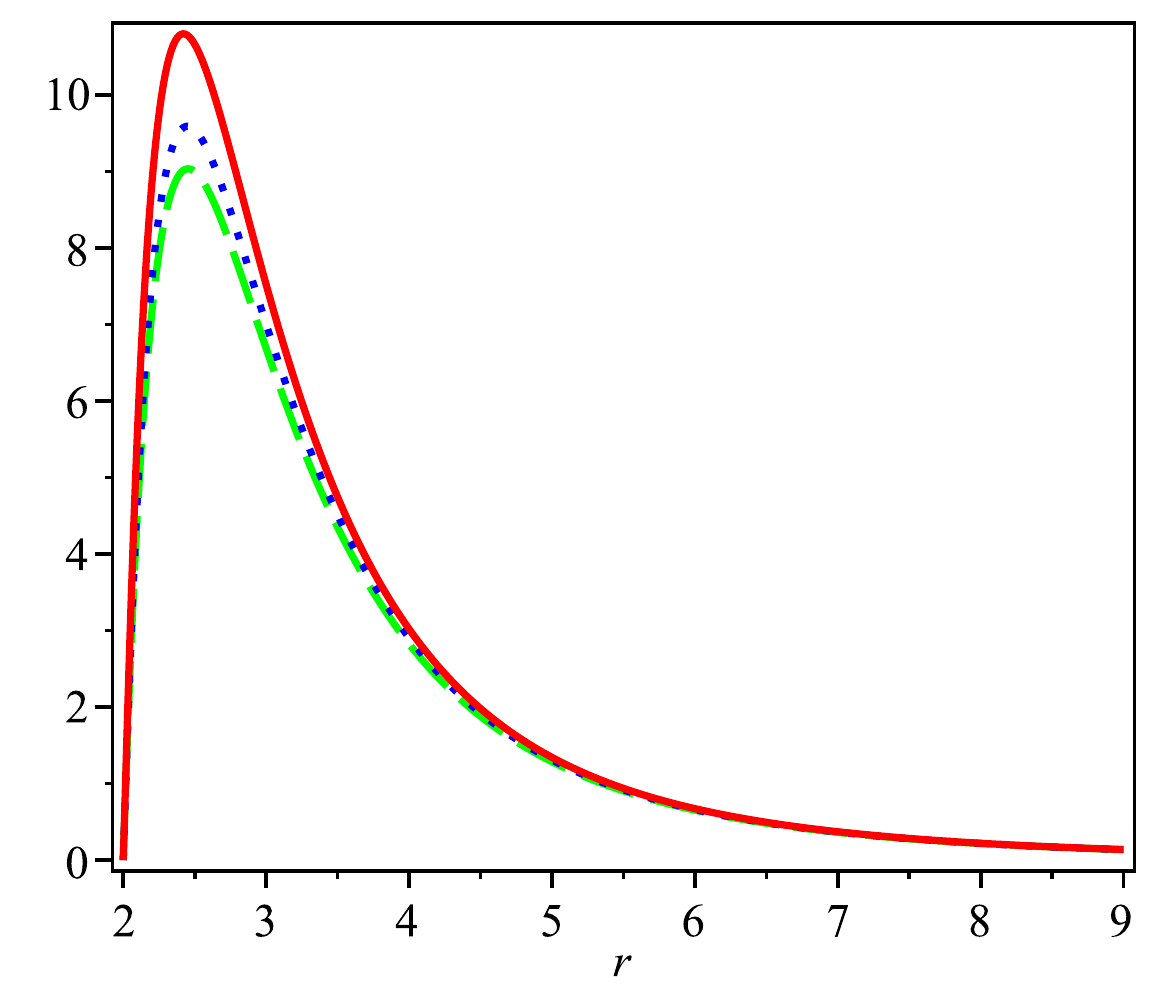}
		\includegraphics[width=0.48\textwidth,height=0.3\textheight]{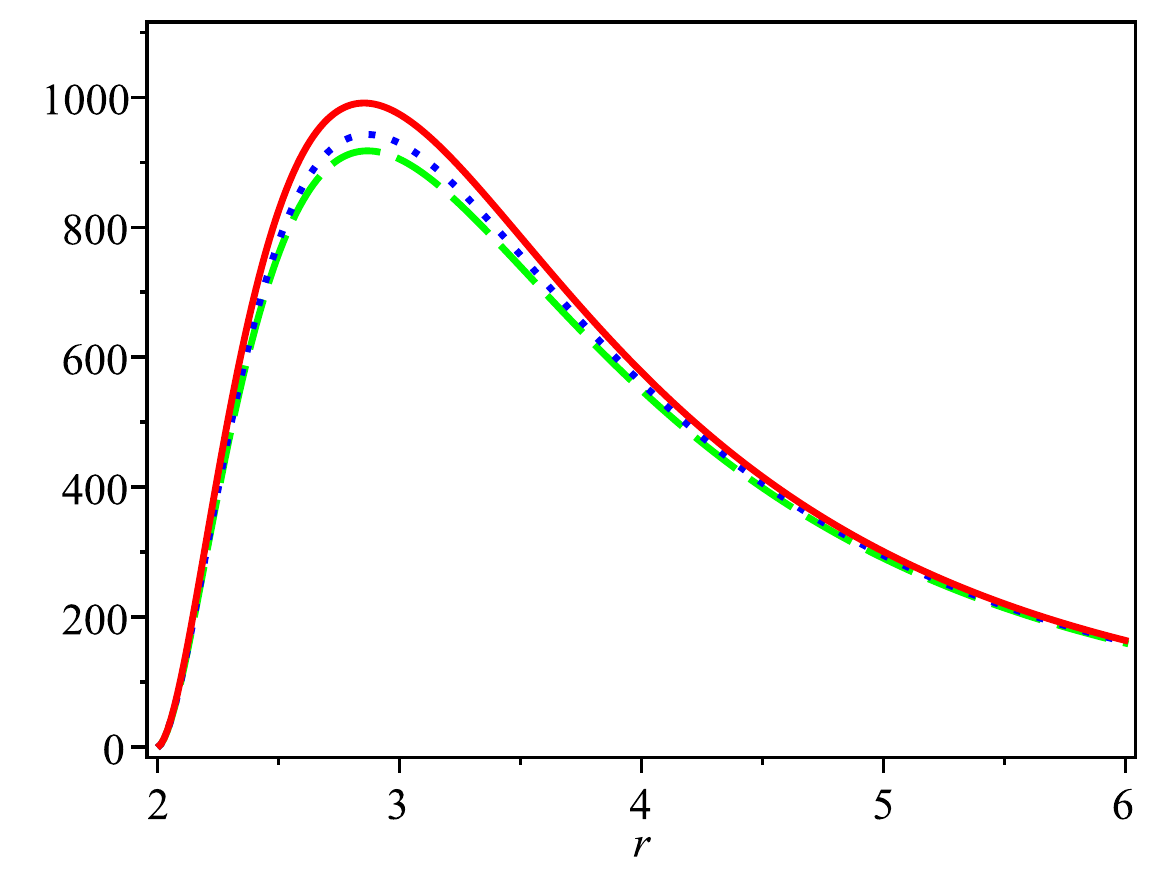}
	\end{center}
	\caption{The behavior of  $\protect\rho  +p_{r}+3p_{t}$ versus $r$ for  $\lambda=-0.0001$ and $\lambda=23.99$ in the left and right panels, respectively. The model parameters have been set as $\alpha=-1.6$ (solid curve), $\alpha=-1$ (dotted curve), $\alpha=-0.5$ (dashed curve), $m=5$, $k=1$ and $r_0=2$ in 5D.}
	\label{wor9phisec}\label{wor6aphisec}\label{worm7phisec}
\end{figure}

\begin{figure}[h]
	\begin{center}
		\includegraphics[width=0.48\textwidth,height=0.3\textheight]{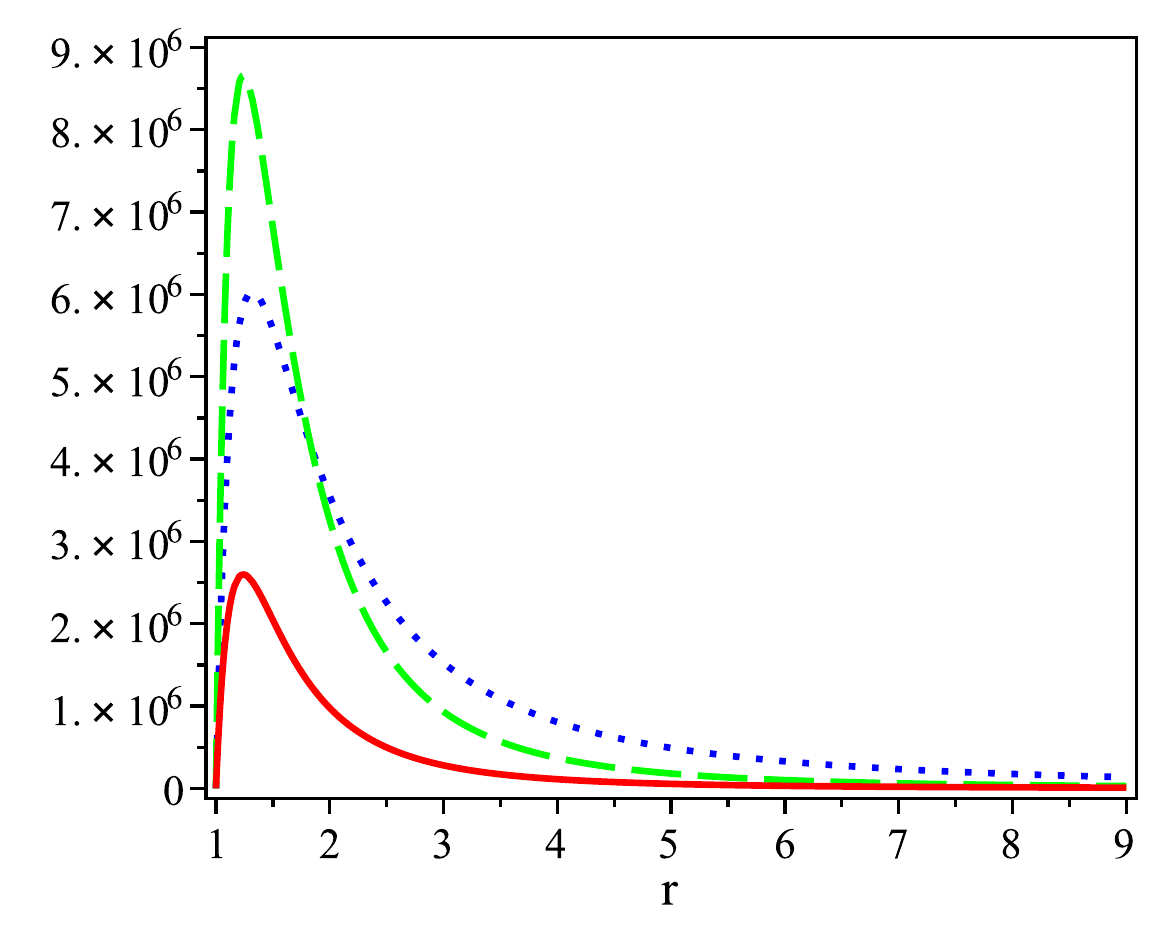}
		\includegraphics[width=0.48\textwidth,height=0.3\textheight]{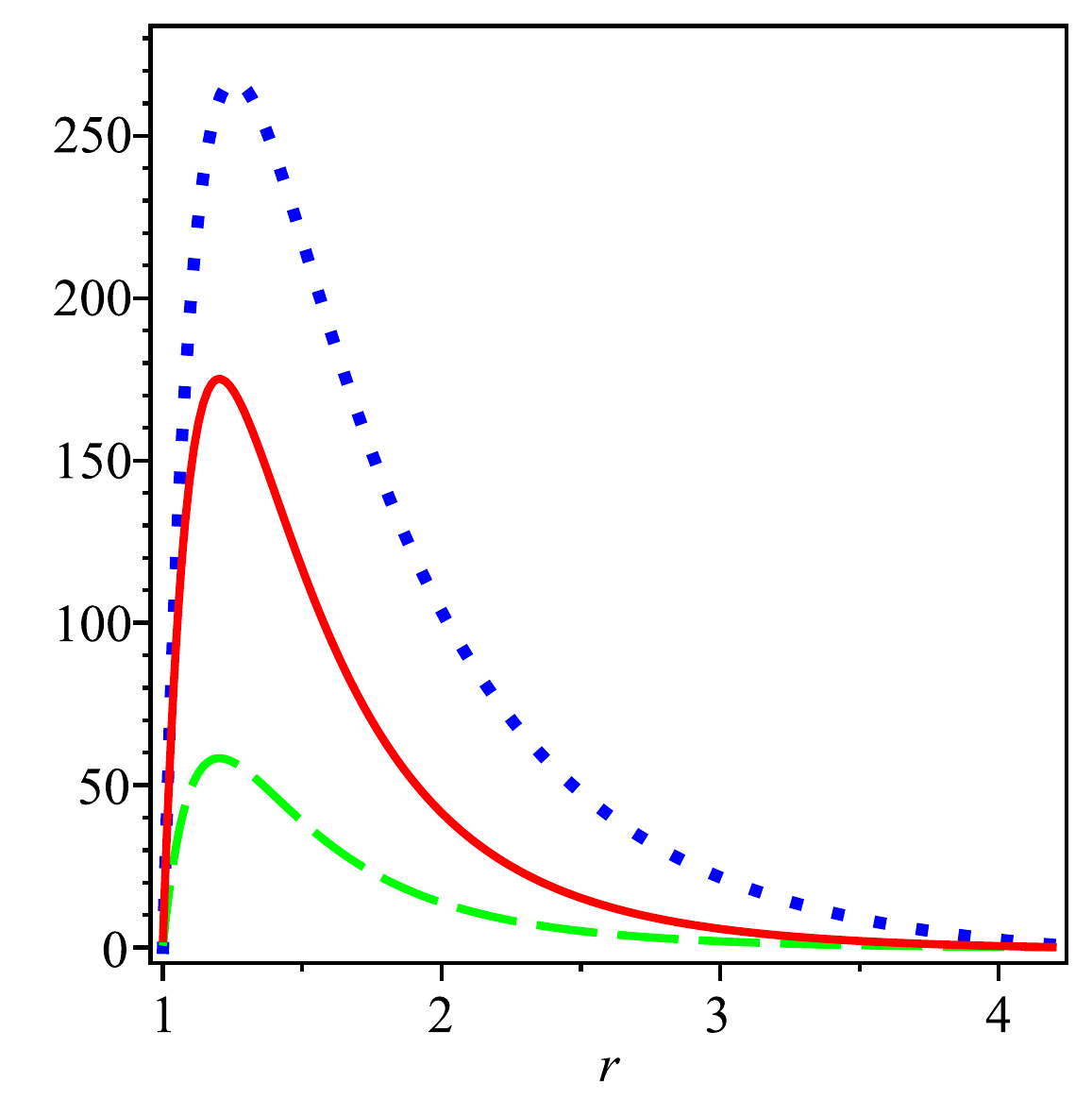}
	\end{center}
	\caption{The behavior of $\protect\rho  +p_{r}+3p_{t} $ (solid), $\protect\rho +p_{r}$ (dotted) and $\protect\rho +p_{t}$ (dashed) versus $r$. The model parameters have been set as $\alpha=0.002$, $m=1$, $\lambda=-2.5$ and $r_0=1$ in the left panel, and for the right panel we have set $\alpha=1$, $\lambda=1$, $m=1$ and $r_0=1$ in 5D.}
	\label{wecsecphinon}
\end{figure}

{The stability and traversability of a wormhole configuration can be significantly improved by using a non-constant redshift function. For instance, Ref. \cite{Hassan:2024} discusses the use of a power-law redshift function within the four-dimensional Gauss-Bonnet framework to construct stable wormholes in Galactic Halos}. Here, we are interested to study the WEC and SEC considering a non-zero redshift function. For this purpose, we choose an asymptotically flat redshift function given by~\cite{msmf}
\begin{equation}
	\phi(r)=\phi_0 \left(\frac{r_0}{r}\right)^{k},~~~~~~~~~k>0.\label{nonphi}
\end{equation}
{This form of redshift function has been also employed in studying wormhole solutions e.g., in $f(Q,T)$ gravity~\cite{fqtred}, wormholes supported by three-form fields~\cite{lobo3f} and wormhole configurations in hybrid metric-Palatini gravity~\cite{salhybpala}, see also~\cite{hendiku} for a comprehensive review.} Now, we investigate the behavior of the energy conditions substituting redshift function Eq.~(\ref{nonphi}) into Eqs.~(\ref{wecc1})-(\ref{wec3}) for the wormhole solutions presented. For case of asymptotically flat solutions the energy conditions are plotted in Figs.~(\ref{wor9phi}), (\ref{wor9phinn}) and (\ref{wor9phisec}). In Fig.~\ref{wor9phi}, one can choose suitable values for the model parameters so that the WEC is satisfied in $r>r_0$ for $\lambda>0$. Also, we have normal matter with choosing negative GB coupling constant, and the GB term has low effects on regions far from the throat. In Fig.~(\ref{wor9phinn}), we show that one can choose {suitable values for the model parameters so that the NEC is satisfie}d in $r>r_0$ for $\lambda<0$. In Fig.~(\ref{wor9phisec}), we show the behavior of SEC for both negative and positive values of $\lambda$ parameter. For the case of asymptotically non-flat solutions, as shown in left panel of Fig.~(\ref{wecsecphinon}) one can consider the model parameters in such a way that the SEC is fulfilled in $r>r_0$. In the right panel of Fig.~(\ref{wecsecphinon}), we show that assuming reasonable values for the model parameters the SEC and WEC are satisfied in the region $r_0<r<r_{+}$ for a de-Sitter wormhole. It may be also possible to find de-Sitter and non-flat wormhole solutions that satisfy the WEC and SEC by choosing positive GB coupling constant in the vicinity of the throat and in all spacetime. 

\section{Equilibrium Conditions}\label{Eqcon}
\begin{figure}
	\begin{center}
		\includegraphics[width=7cm]{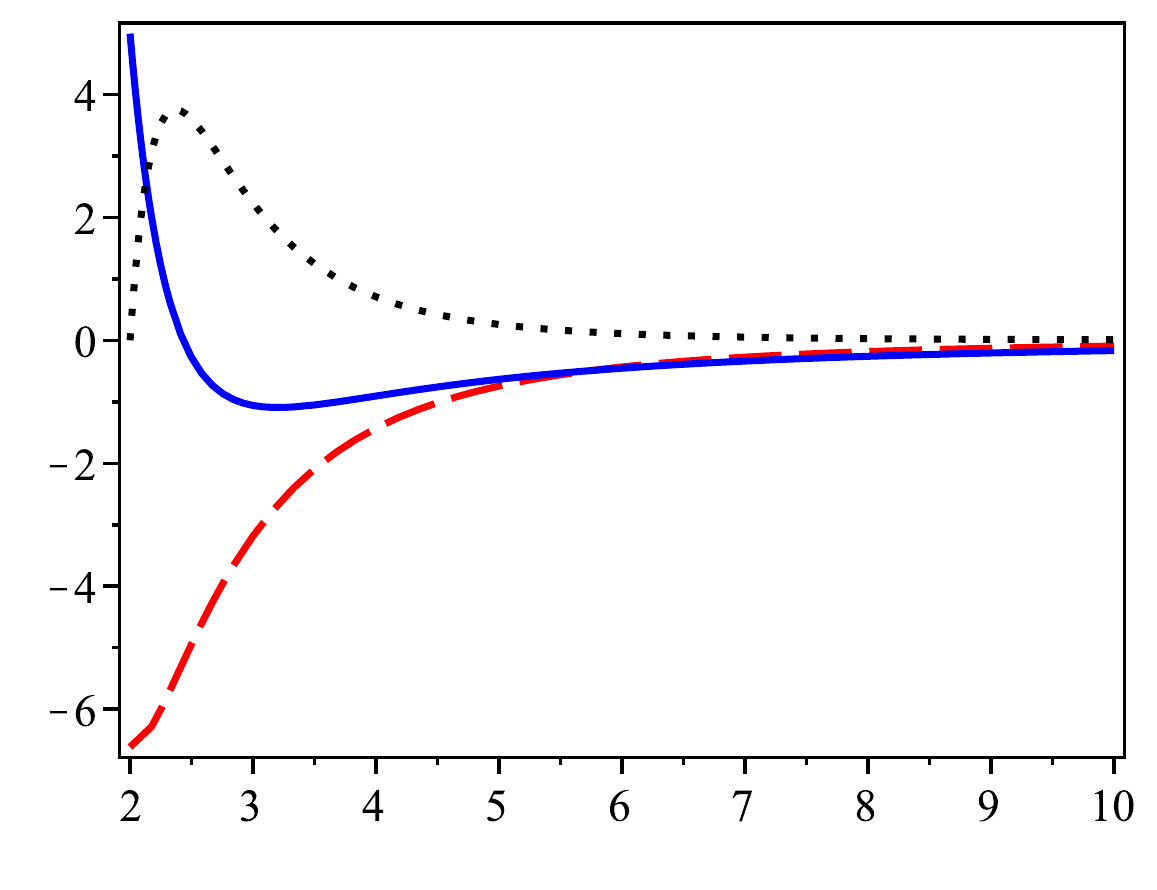}
		\includegraphics[width=7cm]{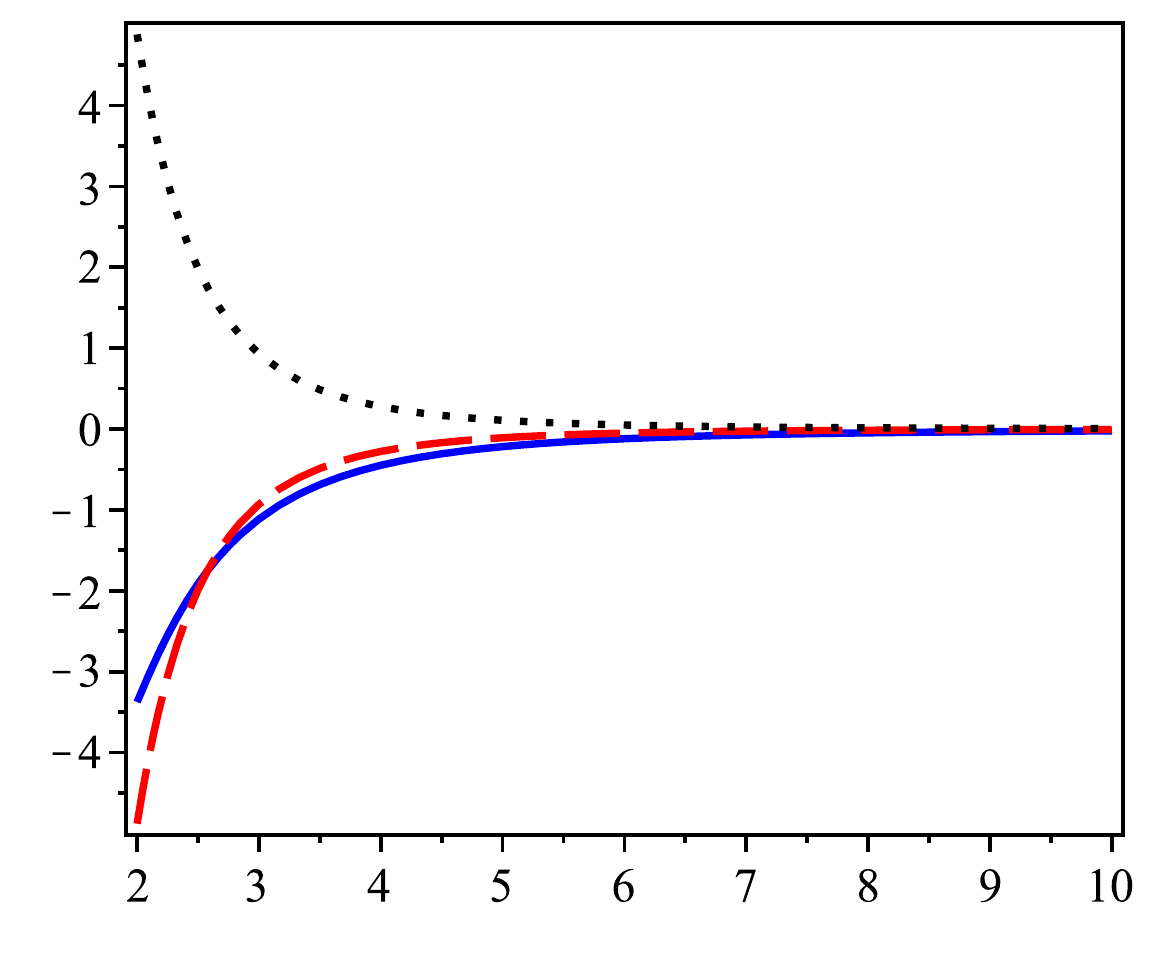}
		\caption {The behavior of $F_h$ (solid curve), $F_a$ (dashed curve) and $F_g$ (dotted curve) against $r$ for asymptotically flat (left panel) and asymptotically non-flat (right panel) solutions. The model parameters have been set as of the left panels in Figs.~\ref{wor9phisec} and~\ref{wecsecphinon}.}\label{figw}
	\end{center}
\end{figure}

In the present section we examine the stability of the obtained wormhole solutions utilizing the equilibrium condition. In the context of GR, this condition can be deduced using the well-known Tolman-Oppenheimer-Volkov (TOV) equation~\cite{rakp}. For the EMT given in Eq.~(\ref{emtt}) the TOV equation is found as
\begin{align}
	-{\frac{dp_r}{dr}} +(n-2)\,{\frac{({\it P_t}-P_r)}{r}}-\phi'(r)(\rho+P_r)=0.
\end{align}
Given the above equation one can investigate the equilibrium state of a wormhole structure by considering the gravitational ($F_g$), hydrostatic ($F_h$) as well as the anisotropic ($F_a$) forces. These forces are defined through the following relations
\begin{eqnarray}
	F_g=-{\phi^\prime(r)}(\rho+p_r),~~~~~~~~~~F_h=-{\frac{dp_r}{dr}},~~~~~~~~~F_a=\frac{n-2}{r}(p_t-p_r).\label{3forces}
\end{eqnarray}
In terms of the above forces the TOV equation can be rewritten as
\begin{align}
	F_g+F_h+F_a=0.\label{stability2}
\end{align}
For the obtained solutions we therefore get the following relations
\bea
F_a &=& \f{(n-2)}{r^5}\left(-r+b\right)\left({r}^{3}+2\alpha b\right)\big((\phi^{\prime})^{2}+\phi ^{\prime \prime}\big) +\f{(n-2)\big[\left(6{\alpha}b r+{r}^{2} \left( {r}^{2}-4{\alpha} \right)\right) r \phi^{\prime}b^{\prime}\big]}{2 r^7}\notag \\
&&+ \frac{(n-2) \phi^{\prime} \big[-18\alpha b^2+ \left(-3\,{r}^{3}+16r{\alpha}\right)b+2{r}^{4}\big]}{2 r^6}\nn&+&\frac{(n-2)\left(b^{\prime}r-3b\right)  \left[2\left(n-5\right) {\alpha}\,b+{r}^{3} \left( n-3 \right) \right]}{2r^7},
\eea
\bea
F_h &=& -\frac{n-2}{2r^7}\bigg[-2r^2 \left( r-b \right)  \left( {r}^{3}+2{\alpha}b\right)\phi^{\prime\prime} +2r\Big(\left(4\alpha b r+r^2 \left( -2{ \alpha}+{r}^{2} \right)  \right)b^\prime-10{\alpha}b ^{2}\notag\\&+& \left( 8r{ \alpha}-2\,{r}^{3} \right) b
\left( r \right) +{r}^{4} \Big) \bigg]+ \frac{(n-2)\big[\left(b^{\prime} r -3 b\right)\big(2\left( n-5 \right){\alpha}b+{r}^{3}	\left( n-3 \right)\big)	\big]}{2 r^7},
\eea
\bea
F_g=\frac{(n-2) \phi^{\prime}\big(r^3+2\alpha b\big)\big[2r \phi^{\prime}(r-b)-b(r)+r b^{\prime}\big]}{2 r^6}.
\eea
Figure (\ref{figw}) shows the graph of gravitational, hydrostatic and anisotropic forces defined in Eq.~(\ref{stability2}) for asymptotically flat (left panel) or asymptotically non-flat (right panel) solutions. It is therefore seen that these forces cancel the effects of each other leaving thus a stable wormhole configuration.
\section{Particle Trajectories Around the Wormhole}\label{parti}
\begin{figure}
	\begin{center}
		\includegraphics[scale=0.348]{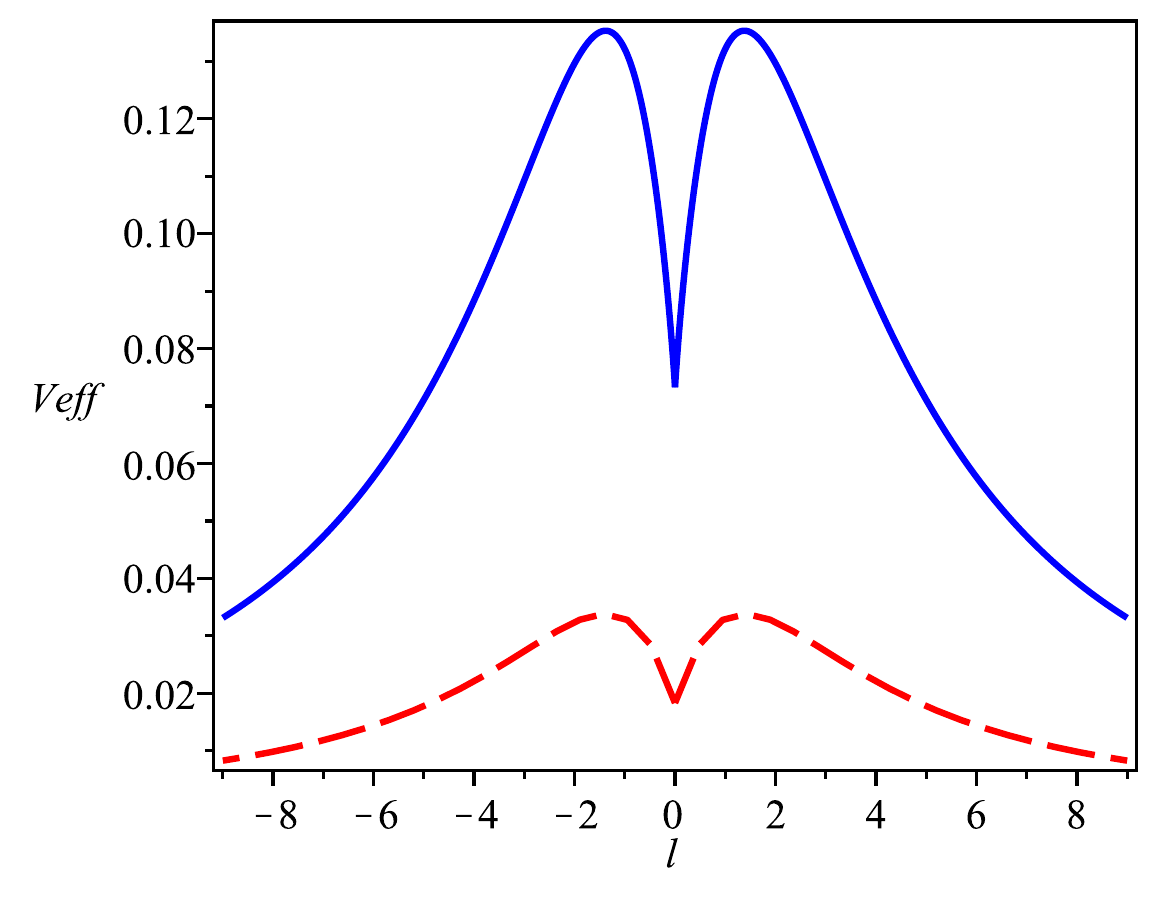}\label{woa}
		\includegraphics[scale=0.368]{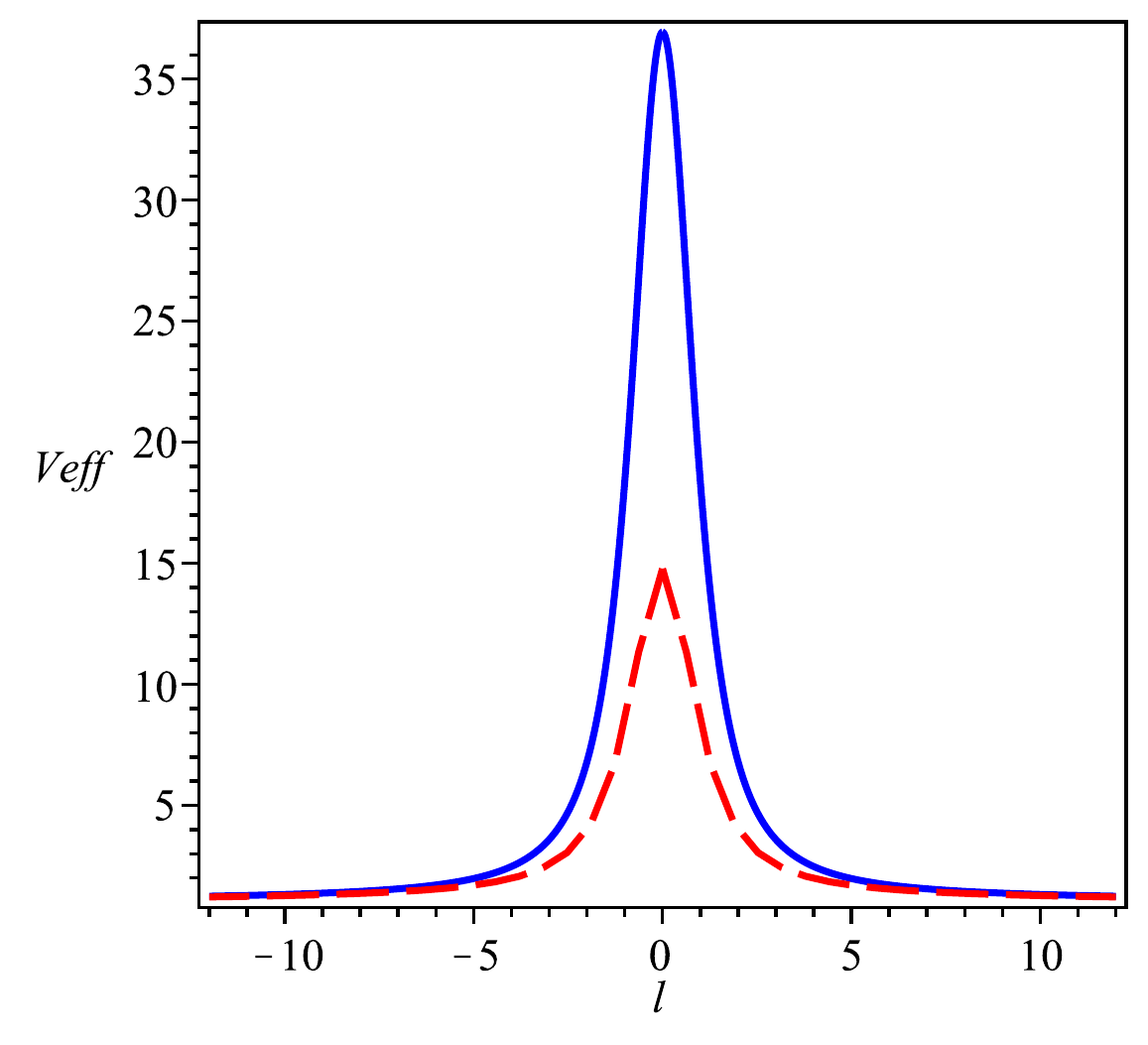}\label{wob}
		\caption{Effective potential $V_{{\rm eff}}$ for null geodesics in left panel with $\phi_0=-2$ and in right panel with $\phi_0=-1$. In these plots we have set the model parameters as $L=2$ (solid curve), $L=1$ (dashed curve) and $\alpha=-0.5$, $\lambda=-10^{-4}$, $n=m=5$, $r_0=1$, $k=1$.}\label{veffw}
	\end{center}
\end{figure}

\begin{figure}
	\begin{center}
		\includegraphics[scale=0.348]{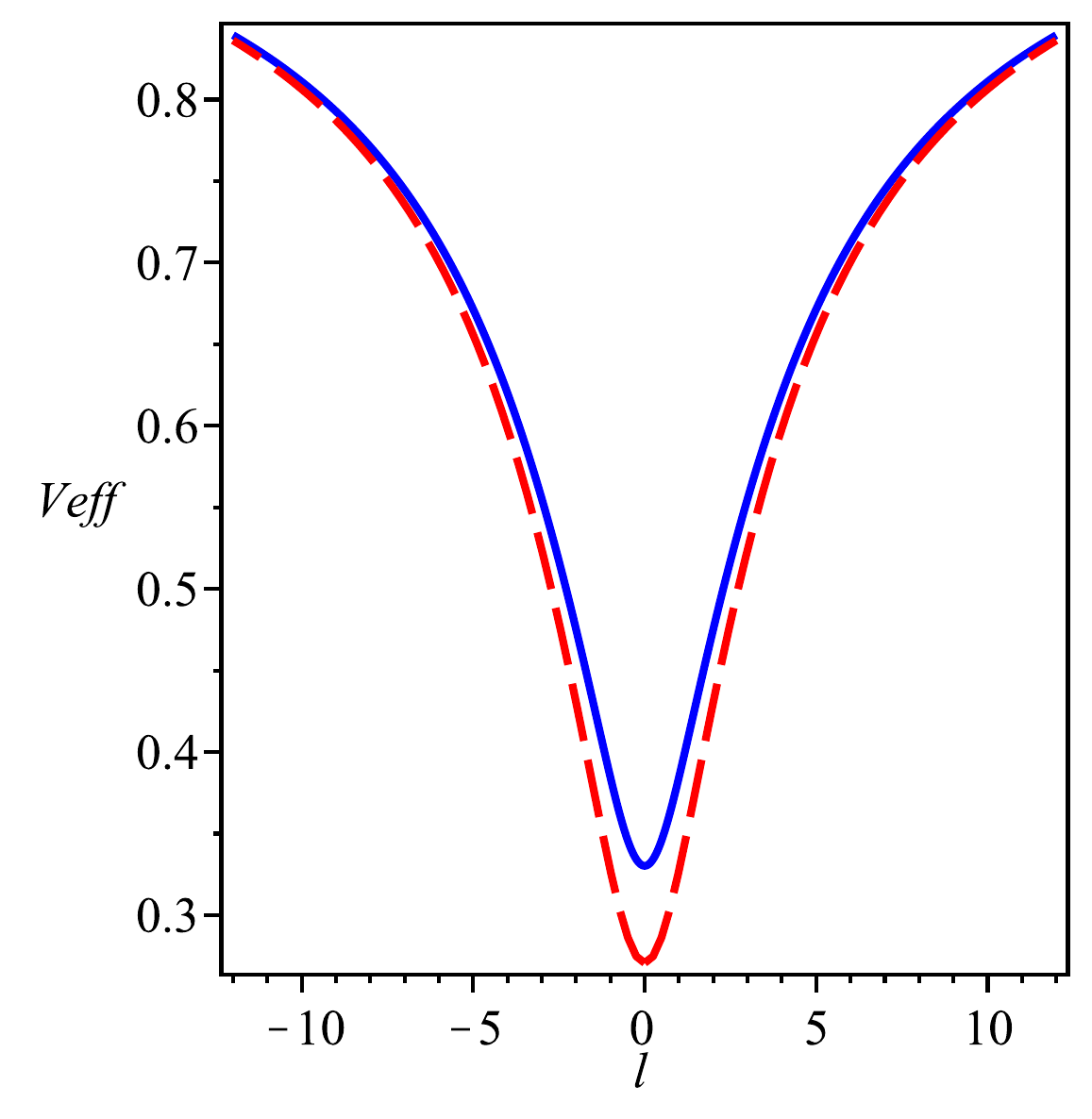}
		\includegraphics[scale=0.368]{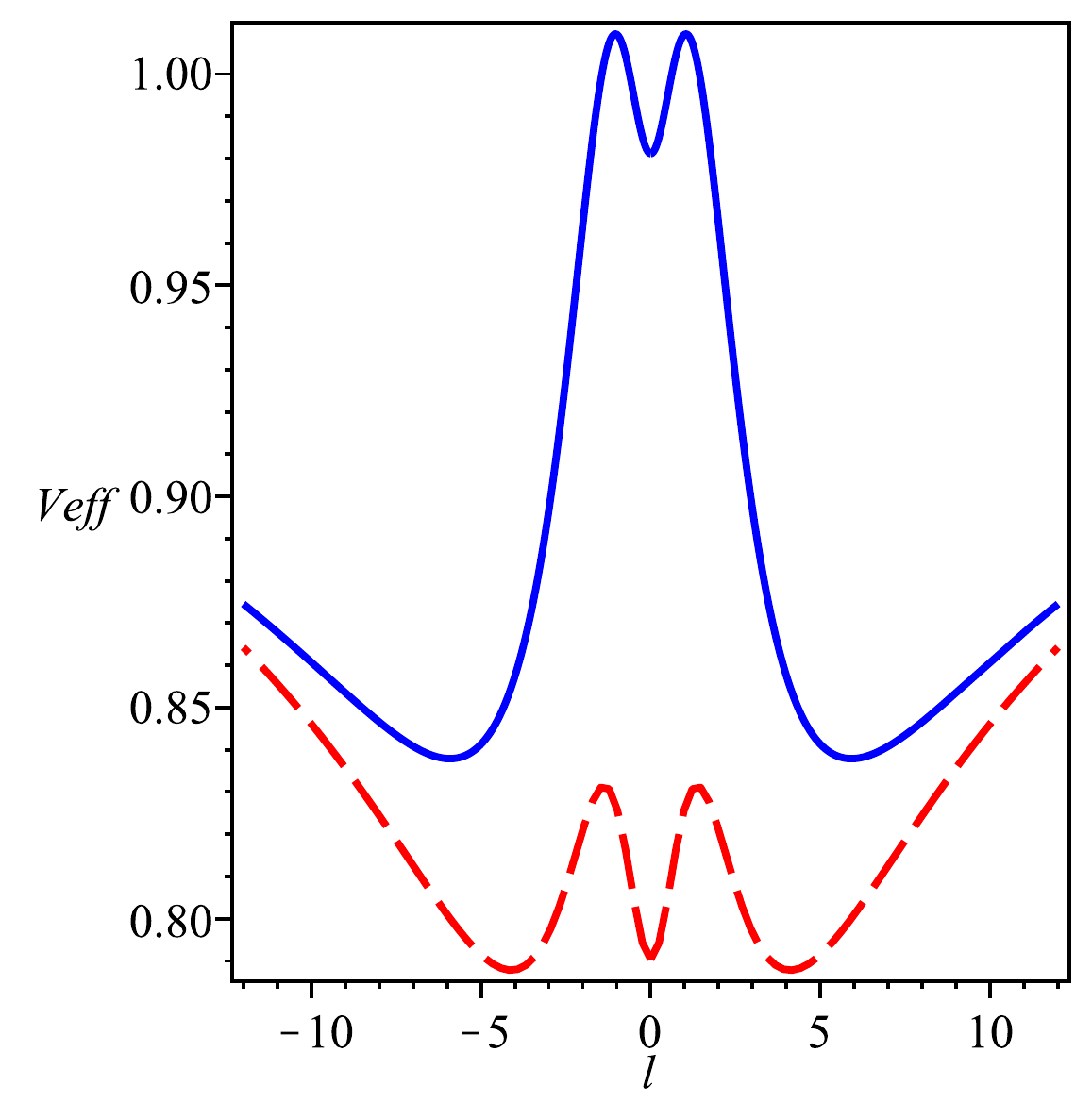}
		\caption{Effective potential $V_{{\rm eff}}$ for timelike geodesics, for $L=1$ (dashed curve), $L=1.2$ (solid curve) in left panel and $L=2.2$ (dashed curve), $L=2.5$ (solid curve) in right panel. The model parameters have been set as $k=1$, $\phi_0=-1$, $\alpha=-1$, $\lambda=-10^{-4}$, $n=m=5$ and $r_0=1$}.\label{veffw2}
	\end{center}
\end{figure}

In this section we analyze geodesic equations in wormhole spacetime described by the metric Eq. (\ref{metr1}), using the Lagrangian formalism~\cite{lagf}. Due to the spherical symmetry of the space–time, the motion of a test particle can be restricted to the equatorial
plane defined by  $\theta_i=\pi/2$, $i>1$. The corresponding Lagrangian for metric (\ref{metr1}) is then found as
\begin{equation}
	\mathfrak{ L} = g_{\mu\nu} \dot{x}^\mu \dot{x}^\nu= -e^{2\phi(r)}\dot{t}^2+\frac{\dot{r}^2}{1-\frac{b(r)}{r}}+r^2\dot{\varphi}^2,
	\label{lag}
\end{equation}    
where a dot denotes derivative with respect to the affine parameter $\eta$. As the Lagrangian is constant along a geodesic we can consider $\mathfrak{L}(x^{\mu} , \dot{x}^\mu)={\epsilon}$ so that time-like and null geodesics correspond to ${\epsilon}=-1$ and ${\epsilon}=0$, respectively. Using the Euller-Lagrange equation, 
\begin{equation}\label{lag2}
\frac{d}{d\eta} \frac{\partial{\mathfrak L}}{\partial\dot{x}^{\mu}}-\frac{\partial{\mathfrak L}}{\partial x^{\mu}}=0,
\end{equation}
one can readily identify the following constants of motion
\begin{equation}\label{lag5}
	\dot{t}=\frac{E}{e^{2\phi(r)}} \  \ {\rm and} \  \ r^2\dot{\varphi}=L,
\end{equation}
where $E$ is the energy and $L$ is the angular momentum of the test particle confined to the equatorial plane.
Inserting these constants of motion into (\ref{lag}) we get
\begin{align}\label{radc}
	\dot{r}^{2}&=\bigg(1-\frac{b(r)}{r}\bigg)\Big(\frac{E^2}{e^{2\phi(r)}}-\frac{L^{2}}{r^{2}}+\epsilon\Big).
\end{align}
It is convenient to rewrite Eq. (\ref{radc}) in terms of the proper radial distance
\begin{equation}\label{mt2}
	l(r)=\pm\int_{r_0}^r\frac{dr}{(1-b(r)/r)^{1/2}},
\end{equation}
which is finite for all finite values of $r$. Note that the spacetime is extended in such a way that $l$ monotonically increases from
$-\infty$ to $+\infty$. $l<0$ or $l>0$ correspond to two parallel universes joined together via a throat at $l=0$. Using the proper radial distance, Eq. (\ref{radc}) takes the simple form 
\begin{align}
	\label{eq:geodesics}
	\dot{l}^{2}=e^{-2\phi(r)}[E^{2}-V_{{\rm eff}}(L,l)],
\end{align}
where the effective  potential is defined as
\begin{align}
	\label{eq:potential}
	V_{{\rm eff}}(L,l)=e^{2\phi\left(r(l)\right)}\left[\frac{L^{2}}{r(l)^{2}}-\epsilon\right].
\end{align}
In what follows, we discuss the trajectory of particles around the wormhole, using the above form for the effective potential. In fact, geodesic equation (\ref{eq:geodesics}) can be interpreted as a classical scattering problem with a potential barrier $V_{{\rm eff}}(L,l)$. Moreover, using Eq. (\ref{lag5}) we can rewrite Eq. (\ref{eq:geodesics}) as an ordinary differential equation for orbital motion 
\begin{equation}\label{lag7}
	\left(\frac{dl}{d\varphi}\right)^2=\frac{{\dot{l}}^2}{\dot{{\varphi}}^2}=\frac{{r(l)}^4}{e^{2\phi\left(r(l)\right)}L^2}\left[E^2- V_{{\rm eff}}(L,l)\right].
\end{equation} 
We note that, in traversable wormhole spacetimes, particles can travel through the throat of the wormhole from one asymptotically flat part of the manifold to the other one. Then, a geodesic can pass through the throat into the other universe if $E^{2}>V_{{\rm eff}}(L,0)$. Similarly, for a geodesic reflected back on the same universe by the potential barrier, we have $ E^{2}<V_{{\rm eff}}(L,0)$. In this case, there is a turning point at $l=l_{{\rm tu}}$ which is obtained by solving $E^{2}=V_{{\rm eff}}(L,l_{{\rm tu}})$.
It is then easy to verify that
\begin{align}
	\frac{dV_{{\rm eff}}}{dl}=2e^{2\phi(r)}\sqrt{1-\frac{b(r)}{r}}\bigg[\frac{\left(r L^2-r^3 \epsilon\right) \phi^{\prime}(r)-L^2}{r^3}\bigg],
\end{align}
and
\begin{eqnarray}
	\frac{d^{2}V_{{\rm eff}}}{dl^{2}}&=&\frac{4e^{2\phi(r)}(r-b)\left(L^2-\epsilon r^2\right){\phi^{\prime}}^2}{r^3}+\frac{2e^{2\phi(r)}(r-b)\left(L^2-\epsilon r^2\right){\phi^{\prime \prime}}}{r^3}\notag \\
	&&-\frac{e^{2\phi(r)}\big[\left(L^2-\epsilon r^2\right)r b^{\prime}+\left(\epsilon r^2-9L^2\right)b+8 r L^2\big]\phi^{\prime}}{r^4}+\frac{L^2\big(r b^{\prime}-7 b +6 r\big)e^{2\phi(r)}}{r^5}.
\end{eqnarray}
A generic feature of this effective potential for the case of zero tidal force is that it possesses a global maximum at the throat
\begin{align}
	\frac{dV_{{\rm eff}}}{dl}\Big|_{l=0}=0,\quad\quad \quad \frac{d^{2}V_{{\rm eff}}}{dl^{2}}\Big|_{l=0}=\frac{L^2(b^{\prime}(r_0)-1)}{r^4}.
\end{align}
The flaring out condition leads to $\frac{d^2V_{{\rm eff}}}{d\ell^2}<0$ at the throat. This clearly has an unstable orbit since it occurs at the maximum of
the potential for $E^{2}=V_{{\rm eff}}(L,l_{0})$. We note that these conditions are independent of whether the geodesic is null or timelike.
For nonzero red shift function we get the effective potential and its derivatives as
\bea
V_{{\rm eff}}(L,l)&=&e^{2k\phi_0(\f{r_0}{r})}\left[\frac{L^{2}}{r(l)^{2}}-\epsilon\right],\\\label{pote1}
\frac{dV_{{\rm eff}}}{dl}&=&\frac{-2e^{2k\phi_0(\frac{r_0}{r})} \sqrt{1-\frac{b(r)}{r}}\bigg[L^2 r^k+k \phi_0 r_0^k (L^2-\epsilon r^2) \bigg]}{r^{3+k}},\\\label{potepri1}
\f{d^{2}V_{{\rm eff}}}{dl^{2}}&=&\f{e^{2k\phi_0(\frac{r_0}{r})}\bigg[L^2 r^{k}+k \phi_0 r_0^{k}(L^2-\epsilon r^2)\bigg] b^{\prime}(r) }{r^{4+k}}\notag \\
&& - \frac{e^{2k\phi_0(\frac{r_0}{r})}\bigg[\big[4 \phi_0^2 k^2 (L^2-\epsilon r^2) r_0^{2 k}+\left(2 k (L^2-\epsilon r^2) +11 L^2 -3 \epsilon r^2\right) \big] k \phi_0 (r r_0)^{k}-7 L^2 r^{2 k}\bigg] b(r)}{r^{5+2 k}}\notag \\
&& + \frac{e^{2k\phi_0(\frac{r_0}{r})}\bigg[ 4  k^2 \phi_0^2 r (L^2-\epsilon r^2) r_0^{2 k} +2\big[L^2(5+k) r^{k+1} r_0^{k} k \phi_0-\epsilon(1+k) r^2\big]+6 L^2 r^{2k +1}\bigg]}{r^{5+2 k}}.\label{poteprim2}
\eea
In what follows, we study null ($\epsilon=0$) and timelike geodesics ($\epsilon=-1$) in detail for the class of asymptotically flat wormhole solutions considering, $m=n=5$. For massless particles ($\epsilon=0$) from Eq.~(\ref{potepri1}) we can find two roots that satisfy equation $V_{\rm eff}^{\prime}=0$, these roots are given by $r_1=r_0$ and $r_2=r_0 (-\phi_0 k)^{1/k}$. We therefore see that at the throat $V_{\rm eff}^{\prime\prime}(L,0)<0$ (a local maximum) for $\phi_0>-1/k$. For the case of $k=1$, we obtain $r_2>r_0$ for $\phi_0<-1$ and hence a local minimum at the throat. The right panel of Fig.~(\ref{veffw}) shows the changes in effective potential with respect to angular momentum for $k=1$ and with $\phi_0\geq-1$ (a local maximum) at throat. Taking the constants as $k=1$ and $\phi_0=-2$, in the left panel of Fig.~(\ref{veffw}), we have a local minimum at the throat and a maximum value for the effective potential at $r_2$. Also, we observe that the height of the potential barrier increases with increasing the angular momentum. For timelike geodesics ($\epsilon=-1$), we can apply equation~(\ref{poteprim2}) at throat to obtain a local minimum at the throat for $\phi_0<-{L^2}/{k(L^2+r_0^2)}$. Also, using Eq.~(\ref{potepri1}) for $k=1$, we find the following two roots for $V_{\rm eff}^{\prime}(L,r_c)=0$ as
\begin{align}
	r_{c,+}=\frac{L\left(-L+\sqrt{L^2-4 r_0^2 \phi_0^2}\right)}{2\phi_0 r_0},~~~~~~\qquad r_{c,-}=\frac{-L\left(L+\sqrt{L^2-4 r_0^2 \phi_0^2}\right)}{2\phi_0 r_0}.
\end{align}
One can then check the stability of orbits through calculating $L$ from equation $V_{\rm eff}^{\prime}(L,r_c)=0$ and substituting the result into Eq.~(\ref{poteprim2}) to obtain
\begin{align}
	V_{\rm eff}^{\prime\prime}(L,r_c)=-\frac{4(1-b(r_c)/r_c)\phi_0(r_c +2 \phi_0 r_0)r_0 e^{\frac{2\phi_0 r_0}{r_c}}}{r_c^3(\phi_0 r_0+r_c)}.
\end{align}
Considering these conditions, one can plot the effective potential as a function of the proper distance for timelike geodesics. As we see in Fig.~(\ref{veffw2}), one can choose the model parameters such that the potential admits a minimum at the throat. In left panel, it is shown that for $L^2<4r_0^2 \phi_0^2$ the effective potential has only a minimum at the throat and in the right one, for $L^2>4r_0^2 \phi_0^2$ one obtains two real roots ($r_{c,\pm}$) that correspond to a maximum and a minimum, respectively at and out of the throat.
\section{Concluding Remarks}\label{conclu}
In the present work, we studied traversable wormhole solutions in EGB gravity. These class of exact solutions which are asymptotically flat, non-flat and de-Sitter in higher-dimensions, are supported by positive (repulsive force) and negative (attractive force) forms for the Casimir energy density. We therefore studied the WEC and SEC for the obtained solutions considering both zero and nonzero redshift functions. It was shown that the WEC and SEC are violated in the spacetime for zero tidal force solutions. While, for a non-constant redshift function along with choosing suitable values for the model parameters and negative GB coefficient, one can find asymptotically flat wormhole solutions satisfying WEC and SEC, except near the throat. Also, Asymptotically non-flat and de-Sitter solutions can be obtained that fulfill the energy conditions throughout the spacetime, except near the throat. Moreover, it was found that the energy conditions can be satisfied for non-flat and de-Sitter wormholes for positive GB constant. In \cite{zubire1}, asymptotically flat wormhole solutions in five dimensions are presented. The energy density supporting these solutions is introduced through vacuum quantum fields confined between two parallel plates in five dimensions, which is generalizable to higher dimensions. Here, we introduced the most general n-dimensional wormhole solutions sustained by Casimir energy. We showed that an appropriate choice of the redshift function allows the energy conditions to be met near the wormhole throat, thereby supporting these structures with a reduced amount of exotic matter. In addition, using the TOV equation, stability of the obtained solutions for both asymptotically flat and non-flat cases were studied. {We further note that in the context of GR ($\alpha_2\rightarrow0$) and for $n=m=4$ with Casimir energy density between two parallel plates, the solutions obtained in the case $\eta\neq0$ can be reduced to the shape function presented in~\cite{trigcasworm2}.} Finally, trajectory of timelike as well as null particles in 5-dimension were discussed using the Lagrangian formalism. We therefore observed that the effective potential for null geodesics admits maximum values at or near the throat, depending on model parameters. This implies that photon spheres could exist at or outside the wormhole throat~\cite{sarmat}. In addition, we found that the minimum value of the effective potential can also be obtained at the throat of the wormhole which corresponds to the existence of an anti-photon sphere~\cite{Shaikh2023}. For the case of timelike geodesics we found that depending on the model parameters, the effective potential admits local minima at or near the throat and maximum values outside the throat. Hence, considering the total energy of a timelike particle it may have the following orbits in the wormhole spacetime: $i)$ orbits in which the particle can travel through the throat of wormhole from one Universe to the other one, this happens when the particle's total energy is greater than the maximum of the effective potential. $ii)$ Trajectories in which the particle is reflected back by the potential barrier and remains in the very Universe. This occurs when the total energy of the particle is less that the maximum of the effective potential, see the right panel of Fig.~(\ref{veffw2}). $iii)$ Trajectories in which the particle can move on oscillatory or circular paths subject to bound orbits in one of the upper/lower Universes, see also~\cite{partpath} for further discussions and details.
\textbf{Data Availability Statement:}  The data that support the findings of this study are available from the corresponding author, [AHZ], upon reasonable request.

\end{document}